\documentclass[]{aa}
\let\ACMmaketitle=\maketitle
\renewcommand{\maketitle}{\begingroup\let\footnote=\thanks \ACMmaketitle\endgroup}
\usepackage[english]{babel}
\usepackage[utf8]{inputenc}
\usepackage[T1]{fontenc}
\usepackage{graphicx}
\usepackage{amsmath,amssymb,wasysym}
\usepackage{txfonts}
\usepackage[colorlinks=true,linkcolor=blue,citecolor=blue]{hyperref}
\usepackage{etoolbox,ulem}
\usepackage{ccaption}
\usepackage{threeparttable}
\usepackage{natbib}

\newcommand{\Msun}{$M_{\odot}$}
\newcommand{\Rsun}{$R_{\odot}$}

\newcommand{\kms}{km s$^{-1}$}
\newcommand{\PSN}{PSN~J09132750+7627410}

\newcommand{\Ha} {\mbox{H$\alpha$}\,}

\begin{document} 

\title{The fate of the progenitors of luminous red novae:\\
Infrared detection of LRNe years after the outburst}

\titlerunning{The fate of the progenitors of LRNe}

\author{A. Reguitti\inst{1,2}\fnmsep\thanks{E-mail: andrea.reguitti@inaf.it},
A. Pastorello\inst{1},
G. Valerin\inst{1} 
}

\authorrunning{Reguitti et al.} 

\institute{
INAF – Osservatorio Astronomico di Padova, Vicolo dell'Osservatorio 5, I-35122 Padova, Italy
\and
INAF – Osservatorio Astronomico di Brera, Via E. Bianchi 46, I-23807 Merate (LC), Italy
}

\date{Accepted 16 November 2025. Received April 2025; in original form October 2024}
 
\abstract{
We present late-time optical and infrared (IR) observations of a sample of nine extragalactic luminous red novae (LRNe) discovered in the past three decades. In all of these cases, the LRN survivors fade below the pre-outburst luminosity of the progenitors in the optical region.
However, they remain visible in the near-IR (NIR) and bright in the mid-IR (MIR) domains for years.
We recover AT~1997bs in \textit{Spitzer} images from 2004, and a residual source is visible in HST and JWST NIR images 27 years after the outburst.
The spectral energy distribution (SED) of AT 1997bs is consistent with that of an orange giant star with a photospheric temperature of 3750-4250~K and a radius of 120-150~\Rsun, without a significant circumstellar dust attenuation.
Similarly, the SED of AT 2019zhd after three years is compatible with a red supergiant star with $T_{ph}\sim3100\pm100$~K and $R\sim350\pm50$~\Rsun.
Another LRN, AT~2011kp, is detected by JWST 12.5 years after the outburst. Its SED, with two excesses at 1.8 and 7.7 $\mu$m, can be explained by a cold ($T\sim450$~K) dusty shell composed of amorphous carbon surrounding a cold expanded source, plus emission from the Pa$\alpha$ line.
We constructed the $[3.6]-[4.5]$ colour curves extending up to more than 7 years for six LRNe, which show a similar evolution: The MIR colour is $\sim-0.5$ mag before the optical maximum light, it becomes bluer after around one year, and then it gradually turns to redder colours in the following years before reaching $[3.6]-[4.5]\sim+1.0$ mag 7 years after the outburst.
We also estimated the masses and the temperatures of newly formed dust years after the LRN onset. We find that LRNe produce dust masses of the order of (1-5)$\times10^{-4}$ (and up to 2$\times10^{-3}$)~\Msun\ between 7 and 13 years after the outbursts. 
Finally, we find that the remnants of LRNe detected years or decades after the merger tend to be expanded and cool objects, similar to red supergiant stars.
}


\maketitle

\section{Introduction}\label{introduction}
In recent years, a growing number of transients have been discovered that have absolute magnitudes at maximum between those of the most luminous novae and the faintest supernovae (SNe), i.e. in the interval $-15 \lesssim M_V \lesssim -10$ mag.
Objects within this range of brightness are usually called `intermediate luminosity optical transients' \citep{Berger2009ApJ...699.1850B}. However, to avoid confusion with other classes with a similar name, we refer to them as `interacting gap transients' \citep[IGTs;][]{Pastorello2019NatAs...3..676P, Cai2022Univ....8..493C}. This family comprises a wide range of astrophysical events, including (but not limited to) erratic outbursts of supergiant stars; giant eruptions of luminous blue variables \citep[LBVs;][]{Humphreys1994PASP..106.1025H}; electron-capture SN candidates, sometimes labelled intermediate-luminosity red transients \citep[ILRTs;][]{Botticella2009MNRAS.398.1041B, Cai2021A&A...654A.157C}; and stellar mergers. Some of the above IGT species share observational similarity with true type IIn SNe without being terminal explosions. For this reason, they are collectively named SN impostors \citep{VanDyk2000PASP..112.1532V, Maund2006MNRAS.369..390M}.

Among the IGTs, in this paper we focus on luminous red novae (LRNe; \citealt{Kulkarni2007Natur.447..458K}), which are thought to be the visible outcome of the coalescence of two non-degenerate stars, which occurs after the ejection of a common envelope \citep[e.g.,][]{Tylenda2011A&A...528A.114T, Ivanova2013A&ARv..21...59I, Kochanek2014MNRAS.443.1319K, Pejcha2014ApJ...788...22P, Howitt2020MNRAS.492.3229H, MacLeod2022ApJ...937...96M, Ivanova2013Sci...339..433I, Chen2024ApJ...963L..35C}. 
Numerous 3D simulations have been conducted to study the common envelope ejection dynamics \citep{Nandez2014ApJ...786...39N, Ivanova2016MNRAS.462..362I, Ohlmann2016ApJ...816L...9O, Pejcha2016MNRAS.455.4351P, Chamandy2018MNRAS.480.1898C, Iaconi2019MNRAS.489.3334I, Prust2019MNRAS.486.5809P, Reichardt2019MNRAS.484..631R, Sand2020A&A...644A..60S, GB2024MNRAS.527.9145G, Kirilov2025ApJ...994L..41K}.
The final outcome of such an event can either be the complete coalescence of the two stars \citep[e.g.][]{Kruckow2016A&A...596A..58K, Schneider2024A&A...686A..45S} or the survival of the binary system but on tighter orbits \citep[e.g.][]{Ivanova2016MNRAS.462..362I, Sand2020A&A...644A..60S}.

The LRNe are identified through a few recurrent observational characteristics (see also \citealt{Pastorello2019review} for a review). Firstly, they present a light curve with a double (sometimes even triple; \citealt{Tylenda2013A&A...555A..16T, Kankare2015A&A...581L...4K, Pastorello2019AA...625L...8P, Cai2022A&A...667A...4C}) peak. The first is a short-duration peak ($\sim$1 week) that is typically very luminous and characterised by a blue colour. This first maximum is followed a few weeks to months later by a long-lasting second peak or a plateau-like phase, which is redder and usually fainter than the first one.
While during the early peak the energy source is cooling emission produced by the initially hot ejecta, the longer plateau-like phase is powered by hydrogen recombination (see \citealt{Metzger2017MNRAS.471.3200M, Matsumoto2022ApJ...938....5M}).

In a few cases, the precursors of LRNe are detected in profound pre-outburst archival images from ground-based facilities or space telescopes \citep{Fraser2011ATel.3574....1F, Mauerhan2015MNRAS.447.1922M, Blagorodnova2017, Cai2019A&A...632L...6C, Pastorello2021_20hat, Blagorodnova2021, Cai2022A&A...667A...4C}.
From these detections, it is possible to measure the mass of the progenitor, and a convincing correlation between this mass and the peak luminosity of LRNe has been found \citep{Kochanek2014MNRAS.443.1319K, Blagorodnova2021, Cai2022A&A...667A...4C}.
However, after the second or third peak, LRNe fade beyond the detection threshold in the optical domain, even when using the largest astronomical facilities, leaving questions about the final fate of the progenitors of LRNe after the main outburst and the type of the merger outcome.

If LRNe were terminal events, we would not expect to detect a visible remnant nor a source that has become significantly fainter than the progenitor's magnitude in the deepest images obtained with space telescopes. However, the disappearance of the source in the optical domain is not sufficient alone to prove a terminal stellar explosion. If the merger outcome is engulfed within a dusty shell, this would obscure the emission at the shorter wavelengths, and prevent detection of the merger outcome in the optical domain. As a consequence, the spectral energy distribution (SED) of the progenitor after the outbursts may shift towards longer wavelengths in the near-infrared (NIR) or even in the mid-infrared (MIR) domains. 
Indeed, it has been observed that dust can form in LRNe at late stages once the dense ejecta cools below the temperature for solid condensation \citep[e.g.][]{Wisniewski2008ApJ...683L.171W, Nicholls2013MNRAS.431L..33N, Banerjee2015ApJ...814..109B, Kaminski2015A&A...580A..34K, MacLeod2017ApJ...835..282M, Iaconi2020MNRAS.497.3166I, Blagorodnova2020, MacLeod2022ApJ...937...96M}.
Therefore, a post-outburst inspection in a very wide range of wavelengths is necessary to securely unveil the fate of an LRN. 

In the past decades and centuries, a few LRNe have been observed in the Milky Way: CK~Vul\footnote{but see \cite{Eyres2018MNRAS.481.4931E, Tylenda2024A&A...685A..49T}} in 1670 \citep{Shara1985ApJ...294..271S, Kato2003A&A...399..695K, Kankare2015A&A...581L...4K}, V4332~Sgr \citep{Martini1999AJ....118.1034M, Tylenda2005_V4332}, OGLE-2002-BLG360 \citep{Tylenda2013A&A...555A..16T}, V838~Mon \citep{Munari2002A&A...389L..51M, Bond2003Natur.422..405B, Tylenda2005_V838}, and V1309~Sco \citep{Mason2010A&A...516A.108M, Tylenda2011A&A...528A.114T}. Another proposed Galactic object was V1148~Sgr \citep{Mayall1949AJ.....54R.191M, Bond2022AJ....164...28B}, although its membership to the LRN class is still controversial.
Given their proximity (a few kiloparsecs), the remnants of some of these events were studied in the MIR \citep{Banerjee2006ApJ...644L..57B, McCollum2014AJ....147...11M, Kaminski2018A&A...617A.129K} and even at longer wavelengths (far-infrared and sub-millimetre) by \textit{Herschel} and ALMA \citep[e.g.,][]{Tylenda2016A&A...592A.134T, Kaminski2020A&A...644A..59K, Kaminski2021A&A...655A..32K}.
Studies on these Galactic objects have also revealed strong evidence for bipolar ejecta \citep{Eyres2017MNRAS.467.2684E}, which supports the scenario of the progenitors being in binary systems.
The progenitors of these low-luminosity Galactic LRNe were all low-mass stellar systems, in agreement with the relation between LRN luminosity at maximum and the progenitor mass proposed by \citet{Blagorodnova2021} and later revised by \citet{Cai2022A&A...667A...4C}.

In contrast, poorer constraints on the progenitors and their post-outburst outcomes are available for the intrinsically brightest LRNe, as they are usually observed in more distant galaxies and, with the exception of the Andromeda Galaxy, outside the Local Group. Characterising these sources is a priority because they are believed to arise from very massive stellar systems (up to 50 \Msun, as in the extreme case of AT 2021aess, \citealt{Guidolin2025}).

\cite{Mauerhan2018MNRAS.473.3765M} conducted a search for the remnant of LRN AT~2014ib / SNHunt248 \citep{Kankare2015A&A...581L...4K, Mauerhan2015MNRAS.447.1922M} using data from the main space telescopes. While a post-outburst source was clearly detected, the images they used were acquired only one year after the LRN onset and hence too early to be sure the merger's aftermath settled into a stable state and for the dust to reveal the true stellar outcome.
Similarly, \cite{Szalai2021ApJ...919...17S} published a single epoch of MIR photometry for LRN AT~2017jfs \citep{Pastorello2019AA...625L...8P} obtained on October 2018 (after 10 months). The object was still bright, at 16 mag, but again the time delay from the outburst was still insufficient to provide reasonable constraints on the LRN survivor.

In this work, we present an archival search for optical, NIR, and MIR data of past extra-galactic LRNe in order to constrain the fate of the progenitors and characterise the resulting stellar products.
The structure of the paper is as follows: In Sect. \ref{sect:sample}, we introduce the sample of LRNe and its selection criteria, and we describe the data sources as well as the data analysis techniques. The main results, including the LRN light curves, are reported in Sect. \ref{sect:results}.
In Sect.~\ref{sect:discussion}, we analyse the SEDs of the remnants, determine the masses of the newly formed dust, and provide our conclusions.

\section{The sample and the observations}\label{sect:sample}
\subsection{The sample}
We analyse a sample of nine extragalactic LRNe documented in the literature. Only objects at least five years old (the oldest was discovered in 1997) in 2025 are considered. The fields of our targets were observed by space telescopes, preferentially in the IR domain. Given their extended wavelength coverage with respect to ground-based facilities, the lower sky background and, in some cases, their excellent spatial resolution (as they operate outside Earth's atmosphere), the space telescopes can detect and resolve targets with fainter apparent magnitudes in outer galaxies.
All the objects in our sample are nearby, being located within a distance of 40 Mpc. This increases the probability to detect the quiescent progenitors years before the LRN event and/or to recover the stellar survivors of LRN outbursts a long time later.

The sample also includes transients that were previously classified as SN impostors (AT~2007sv and AT~2015fx), but for which we now offer a LRN reclassification. The final sample is presented in Table \ref{Tab:sample}.
As for some objects an official International Astronomical Union (IAU) name was not available, we take the opportunity offered by this work to give them a proper IAU designation in the Transient Name Server\footnote{\url{https://www.wis-tns.org/}}.

\begin{table*}\centering
\caption{Luminous red nova sample analysed in this paper.}
\label{Tab:sample}
\begin{threeparttable}
\begin{tabular}{llllll}
\hline
Object name & Host galaxy & Distance (Mpc) & $E_{tot}(B-V)$ & MJD of 1st peak & Reference \\
\hline
AT 1997bs   & NGC 3627  & 9.4±0.3   & 0.04 & 50560   & (1) \\ 
AT 2007sv   & UGC 5979  & 18.85±1.0 & 0.83 & <54454\tnote{1} & (2) \\
AT 2011kp\tnote{2} & NGC 4490  & 9.6±1.3   & 0.32 & 55797   & (3) \\ 
AT 2013lw\tnote{2} & UGC 12307 & 39.7±2.7  & 0.22 & 56481\tnote{3}   & (4) \\
AT 2015fx\tnote{2} & NGC 2748  & 23.8±2.0  & 0.024 & 57069   & (5) \\ 
AT 2015dl\tnote{2} & M 101     & 6.4±0.5   & 0.008 & 57070   & (6) \\
AT 2018bwo  & NGC 45    & 6.8±0.5   & 0.02 & <58252\tnote{1}   & (7) \\
AT 2019zhd  & M 31      & 0.78±0.01 & 0.055 & 58892   & (8) \\
AT 2020hat  & NGC 5068  & 5.2±0.2   & 0.37 & 58954   & (9) \\
\hline
\end{tabular}
    \begin{tablenotes}\footnotesize
    \item[1] Discovered after the first peak. The discovery date is used otherwise.
    \item[2] Alternative names: NGC4490-OT2011 (AT 2011kp), UGC 12307-OT2013 (AT 2013lw), \PSN\,(AT 2015fx), M101-2015OT1 (AT 2015dl).
    \item[3] Epoch of the second, red peak.
    \end{tablenotes}
\end{threeparttable}
\tablefoot{References: (1) \cite{Kochanek2012ApJ...758..142K}; (2) \cite{Tartaglia2015MNRAS.447..117T}; (3) \cite{Smith2016}; (4) \cite{Pastorello2019review}; (5) \cite{Tartaglia2016ApJ...823L..23T}; (6) \cite{Blagorodnova2017}; (7) \cite{Pastorello2023}; (8) \cite{Pastorello2021_19zhd}; (9) \cite{Pastorello2021_20hat}.
The total extinction is the sum of the Galactic and host contributions (when known).
}
\end{table*}

For each object, the phases considered in this paper are with respect to the epoch of the first, blue maximum. We also adopted the same distance and total extinction values as reported in their reference articles (see Table \ref{Tab:sample}).

\subsection{The observations}\label{sect:observations}

The imaging data analysed in this paper were obtained with different space facilities, including: 1. the Hubble Space Telescope (HST) with multiple cameras: the Wide Field and Planetary Camera 2 (WFPC2, \citealt{Holtzman1995PASP..107.1065H}), the Wide Field Camera 3 (WFC3), and the Advanced Camera for Surveys (ACS, \citealt{Sirianni2005PASP..117.1049S}); 2. the Spitzer Space Telescope (SST, \citealt{Werner2004ApJS..154....1W, Gehrz2007RScI...78a1302G}) with the Infrared Array Camera (IRAC, \citealt{Fazio2004ApJS..154...10F}); 3. the Wide Infrared Space Explorer (WISE, \citealt{Wright2010AJ....140.1868W}) during the Near-Earth Object WISE (NEOWISE, \citealt{Mainzer2011ApJ...731...53M, Mainzer2014ApJ...792...30M}) mission; 4. the James Webb Space Telescope (JWST, \citealt{Gardner2006SSRv..123..485G}) with the Near Infrared Camera (NIRCam, \citealt{Rieke2023PASP..135b8001R}) and Mid-Infrared Instrument (MIRI, \citealt{Dicken2024A&A...689A...5D}).

We retrieved publicly available HST images from the Mikulski Archive for Space Telescopes (MAST\footnote{https://mast.stsci.edu/portal/Mashup/Clients/Mast/Portal.html}). For SST and WISE, we collected the images from the Spitzer Heritage Archive\footnote{https://irsa.ipac.caltech.edu/applications/Spitzer/SHA/} and the WISE Image Service\footnote{https://irsa.ipac.caltech.edu/applications/wise/}, respectively, both hosted at the NASA/IPAC Infrared Science Archive (IRSA\footnote{https://irsa.ipac.caltech.edu/frontpage/}).
The HST, SST, and JWST data were already fully reduced and flux-calibrated by dedicated processing pipelines (versions 2024.10.2.3, S19.2.0, 1.16.1, respectively), i.e. their counts have been converted into MJy/sr.
For SST, we obtained the level 2 (post-basic calibrated data) images. For HST and JWST, we retrieved the level 3, calibrated images.
The WISE images required a longer reduction procedure.
We retrieved single-exposure frames obtained by the NEOWISE-Reactivation mission - which scanned the entire sky once every six months in the $W1$ (3.4 $\mu$m) and $W2$ (4.6 $\mu$m) filters - from the IRSA Archive. During each of these passages, individual fields can be observed several times within a few days. In these cases, we co-added those taken in this short temporal interval into a single stack image. Finally, we performed Point Spread Function (PSF) fitting photometry from a template-subtracted image. As templates, images obtained a long time (many years) before or after the outburst were used (see Sect.~\ref{sect:results} for details). The WISE magnitudes were calibrated against the WISE All-Sky Data Release catalogue \citep{Cutri2012wise.rept....1C}.

To identify our targets in the high spatial resolutions images (from HST and JWST), we geometrically transformed the pre-outburst images with the detected progenitor to the late-time data. By doing this, we achieved high-precision relative astrometry between the two sets of data. We registered a number of point-like sources in common between the two datasets. Then, using \texttt{geomap} and \texttt{geoxytran} (two \texttt{IRAF} tasks), we carried out a geometrical transformation between the sets of coordinates. 
Later on, for all the objects we performed photometry with the PSF-fitting technique using the \texttt{SNOoPY} pipeline\footnote{{\sl ecsnoopy} is a package for SN photometry using PSF fitting and/or template subtraction developed by E. Cappellaro. A package description can be found at \url{http://sngroup.oapd.inaf.it/snoopy.html}.} v. 3.2.024 - adapted to run on images taken with space telescopes - to obtain the instrumental magnitudes. 
The images from HST and JWST were then calibrated using zero points provided by the images or against a specific catalogue (see the text). The aperture correction calculated by the software was applied, while the local background was estimated by fitting it with a low-order 2D polynomial function, therefore accounting for its non-uniformity.

\section{Results}\label{sect:results}
In this section, all the magnitudes are reported in the AB mag system \citep{Oke1983ApJ...266..713O}. The upper limits are calculated considering a detection threshold of $2.5\sigma$. The photometric tables with the magnitudes for each object are available at the CDS. 
In all figures showing the location of the LRNe analysed in this paper within their host environments, north is up and east is to the left.

\subsection{AT 1997bs}\label{sect:97bs}
This object, known in the past with a SN designation (SN~1997bs), has been extensively studied in the literature \citep{VanDyk1999AJ....118.2331V, VanDyk2000PASP..112.1532V, Li2002PASP..114..403L, Kochanek2012ApJ...758..142K}. The transient was initially considered an SN impostor, given its peak luminosity of only $M_V=-13.8$ mag \citep{VanDyk2000PASP..112.1532V}.
\cite{VanDyk1999AJ....118.2331V} identified a progenitor candidate in HST/WFPC2 images acquired in 1994, with an apparent magnitude of $F606W=22.86\pm0.16$ mag, corresponding to an absolute magnitude of $M_V\sim-7.4$ mag (see also \citealt{Adams2015MNRAS.452.2195A} for an updated estimate of the progenitor's colour). As discussed in \cite{VanDyk2000PASP..112.1532V}, the colour evolution of the transient was peculiar, rapidly evolving towards redder colours, reaching $V-I>3$ mag. 
\cite{Pastorello2019review} noted its light curve was reminiscent of those of LRNe, and proposed a reclassification of the object as a LRN.

\begin{figure}
\includegraphics[width=1\columnwidth]{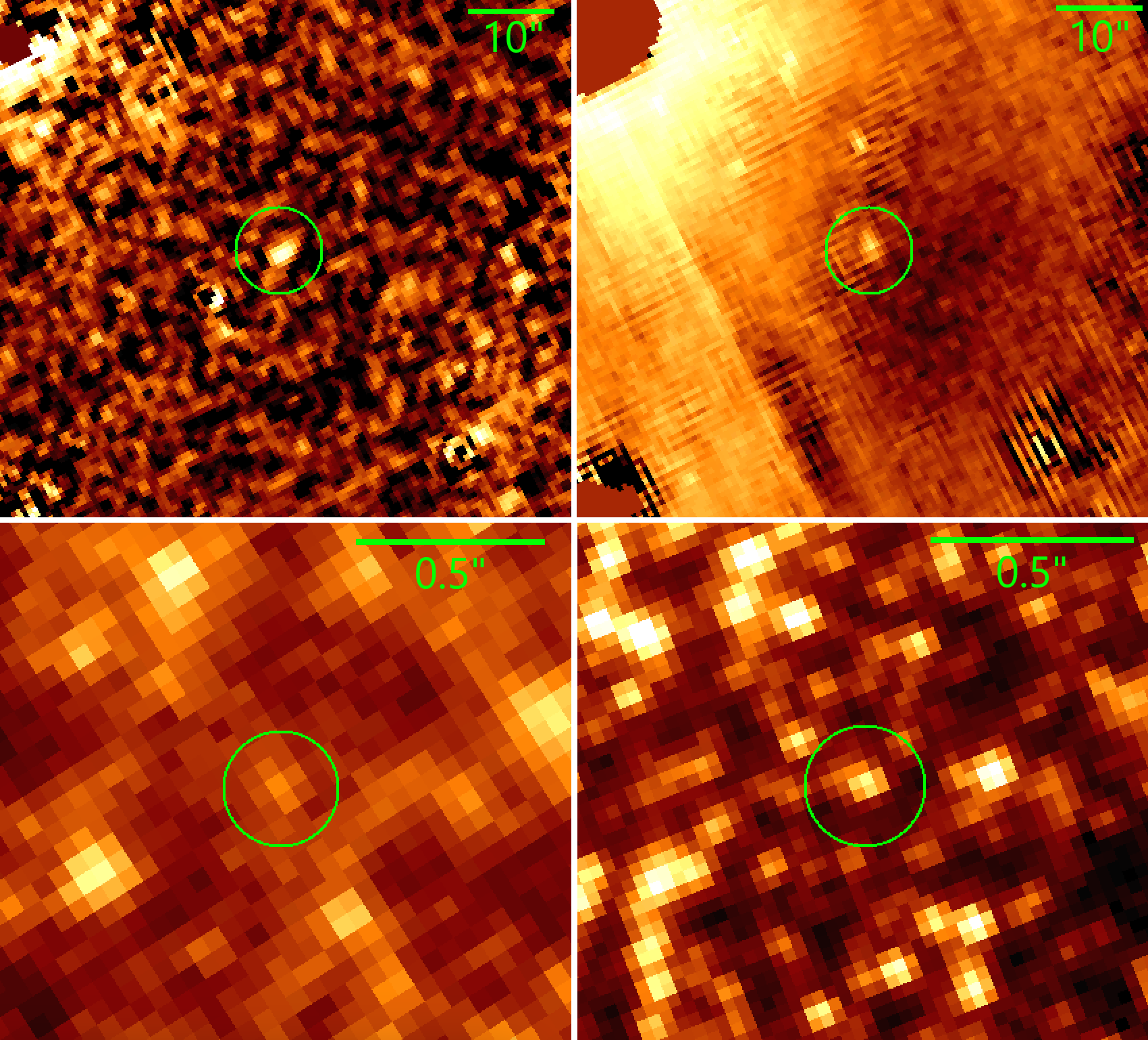}
\caption{Very late time detections of AT 1997bs.
Top-left: Detection of AT 1997bs in an SST $Ch2$ filter image obtained in 2004 (after the subtraction of a template image). Top-right: Same, but for the $Ch4$ filter (using a JWST/MIRI $F770W$ image as a template). The green circles are centred at the transient position and has a radius of $5''$.
Bottom-left: Detection of AT 1997bs in $F814W$ filter in HST/ACS images obtained in 2023.
Bottom-right: Detection of AT 1997bs in $F150W$-filter JWST/NIRCam images obtained in 2023. The green circle are centred at the position of the transient and have a radius of $0.15''$.}
\label{fig:97bs}
\end{figure}

\cite{Kochanek2012ApJ...758..142K} presented archival SST/IRAC images of the field of AT 1997bs taken during the Cold Mission in May 2004 (+7 years), but they only provided upper limits in the $Ch1$ filter (3.6 $\mu$m, indicated also as [3.6]) and $Ch2$ filter (4.5~$\mu$m, or [4.5]). However, from the analysis of the same dataset, \citealt{VanDyk2012ApJ...746..179V} reported the detection of the source in those bands; they provide flux densities of approximately 40 and 50 $\mu$Jy in the $Ch1$ and $Ch2$ filter, corresponding to 19.89 and 19.65 AB mag, respectively (see their Figure 4).
In order to accurately estimate the magnitude of the post-outburst source in 2004 in all IRAC filters\footnote{including $Ch3$ and $Ch4$, indicated as [5.8] and [8.0] because their central wavelengths are expressed in $\mu$m.}, we applied the template-subtraction technique. For the $Ch1$ and $Ch2$ filters, we used very late SST images obtained in September 2019, when no source was detected at the LRN location, as templates.
There is a marginal detection in the $Ch1$ filter image, and a more robust one in the $Ch2$ filter (Fig.~\ref{fig:97bs}, top).
We also attempt to recover information of the source in the SST/IRAC $Ch4$ filter image by subtracting a JWST/MIRI $F770W$ image (see below) as a template. 
For the $Ch3$ filter, the ideal template would be in the JWST/MIRI $F560W$ band, but an image with this filter was not available. Therefore, we attempted to use the JWST $F770W$ image again as a template. As this choice can lead to a wrong estimation of the source flux, we add a conservative systematic error of 0.1 mag. We tentatively identify AT 1997bs in $Ch4$ (Fig.~\ref{fig:97bs}, top right), and the source is also detected in $Ch3$.

After the observations of \cite{Adams2015MNRAS.452.2195A}, AT~1997bs is still barely detectable by HST/ACS in the $F814W$ filter in 2019 (+22 years) and on 2023 May 10 (+26 years). 
In 2023, we measure $F814W=25.5\pm0.3$ mag, which matches the magnitude estimate in 2014 of \cite{Adams2015MNRAS.452.2195A}, $F814W=25.5\pm0.2$ mag. As a consequence, the source has maintained a constant luminosity in the nine years between the two observations.

Deep JWST observations of the host galaxy NGC 3627 conducted in 2023 (+25.7 years) and 2024 (+27.7 years), with both NIRCam and MIRI, reveal that AT 1997bs is still detectable in the $F150W$ (Fig. \ref{fig:97bs}, bottom) and $F200W$ images.
AT 1997bs is barely detectable in the $F277W$ filter, while it has finally faded below the detection threshold in the MIR.

\subsection{AT 2007sv}
AT 2007sv (formerly named SN 2007sv) was classified as an individual major outburst produced by a massive star \citep{Tartaglia2015MNRAS.447..117T}. \cite{Smith2011MNRAS.415..773S} only reported a lower limit on the absolute magnitude of the progenitor of $>-11.7$ mag. However, its light curves and spectra closely resemble those of LRNe, such as AT 1997bs, although the post-peak photometric evolution in the red bands shows a short-duration plateau rather than a broad peak. Though, the presence of a plateau is a quite frequent features in LRN light curves \citep[][and references therein]{Pastorello2023}.
\cite{Tartaglia2015MNRAS.447..117T} measured an outflow velocity of 2000 \kms, which is difficult to reconcile with an LBV wind, whose speeds are usually up to 1000 \kms\,(see \citealt{Smith2017hsn..book..403S}).
Instead, it is consistent with the ejecta velocities observed in some of the brightest LRNe, such as AT 2014ib \citep{Mauerhan2015MNRAS.447.1922M} and AT 2021biy \citep{Cai2022A&A...667A...4C} in the early phases.
In addition, numerous absorption lines from Fe~II, Sc II and Ba II are visible in the spectra.
Based on all of these observables, AT 2007sv matches all the classification criteria of a LRN event.

\begin{figure}
\includegraphics[width=1\columnwidth]{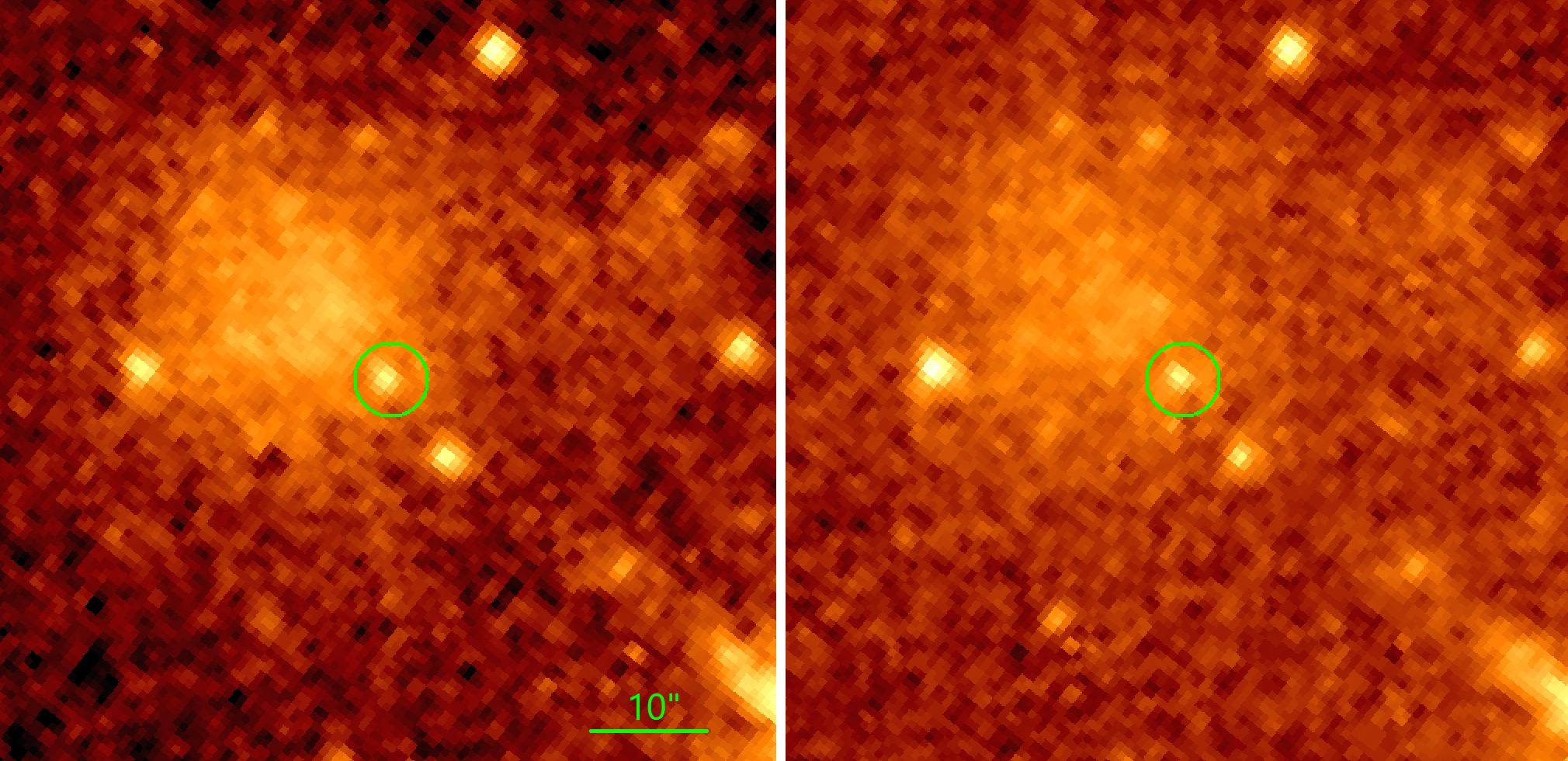}
\caption{Detection of AT 2007sv in 2010 images obtained by SST in the $Ch1$ (left) and $Ch2$ (right) filters. The green circles are centred at the transient position and have a radius of $3''$.}
\label{fig:07sv_Spitzer}
\end{figure}

SST observed the field of AT 2007sv in 2009 and 2010 ($\sim$2 years after discovery), and the transient is clearly detected in the $Ch1$ and $Ch2$ filters (Fig.~\ref{fig:07sv_Spitzer}). 
The source shows a modest brightening between the first and second epoch, by 0.06 mag in $Ch1$ and 0.2 mag in $Ch2$, which is consistent with what has been observed in other LRNe presented in this work at late times.

\subsection{AT 2011kp (NGC 4490-OT2011)}

\begin{figure*}\centering
\includegraphics[width=2.04\columnwidth]{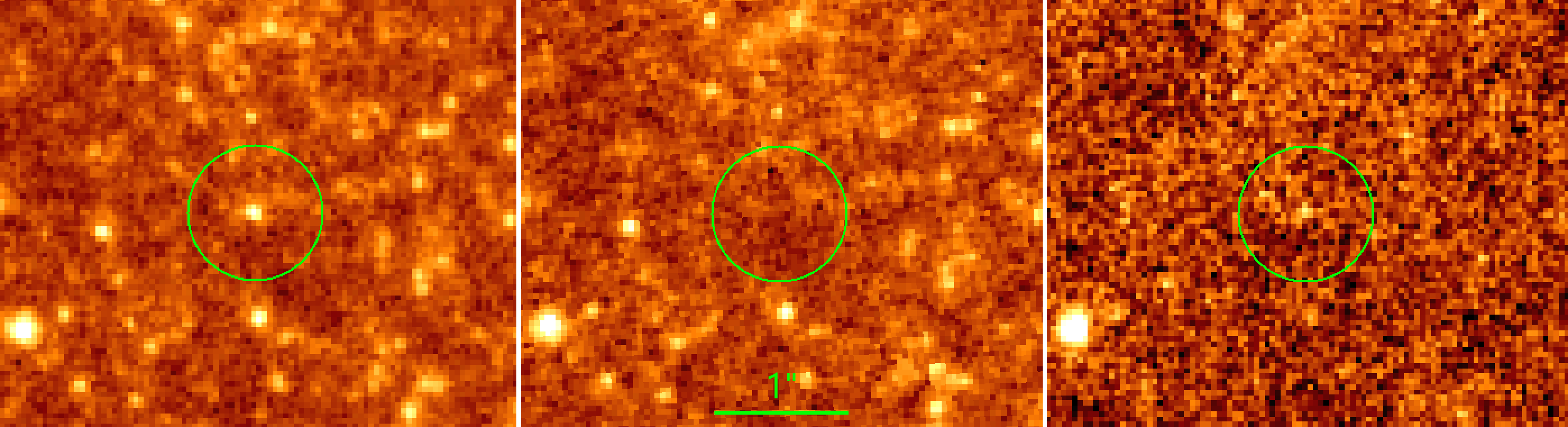}
\caption{Images from HST of the field of AT 2011kp. Left: $F814W$ filter image obtained on 2013 October 30. Centre: $F814W$ filter image taken on 2014 January 12. Right: $F657N$ HST image obtained on 2014 October 23. The green circles are centred at the coordinates of the transient and have a radius of $0.5''$.
}
\label{fig:ngc4490_hst}
\end{figure*}

AT 2011kp\footnote{Alternative designations used in the literature for AT~2011kp are NGC4490-2011OT1 and NGC4490-OT2011.} is a LRN studied by \cite{Smith2016} and \cite{Pastorello2019review}. 
\cite{Fraser2011ATel.3574....1F} found a progenitor candidate in an archival HST image before the outburst, with an absolute magnitude of $F606W = -6.2$ mag. Unfortunately, no colour information was available for this source. However, \cite{Smith2016} inferred an initial mass of 20-30 \Msun\,for the progenitor.

\cite{Smith2016} presented optical (HST) and mid-infrared (SST) observations up to +2.5 years of AT 2011kp, in which the object was still visible. Here, we analyse more recent data obtained after those published in their paper. 
Notably, while the LRN is still detected in the $F814W$ filter in October 2013 (Fig. \ref{fig:ngc4490_hst}, left), only 3 months later (in January 2014), the source suddenly vanishes (Fig.~\ref{fig:ngc4490_hst}, centre). The same result is obtained in August 2014, and in October 2014 with the $F547M$ filter.
However, the source does not disappear in all optical filters given that, at the same epoch, a very faint source is seen at the LRN position in a $F657N$-filter image (centred on the \Ha emission line, Fig. \ref{fig:ngc4490_hst}, right).
This implies that the outcome of the LRN event is still emitting in the \Ha line.

\begin{figure}
\includegraphics[width=1\columnwidth]{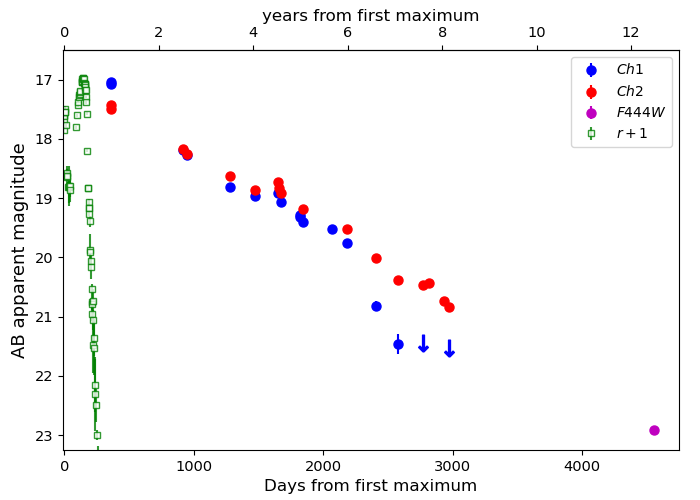}
\caption{Mid-infrared light curve of AT 2011kp spanning 12.5 years of evolution. The 2024 JWST/NIRCam $F444W$ filter measurement is also shown as its bandpass is similar to that of the $Ch2$ filter. Upper limits are marked with downwards arrows. For reference, the $r$-band light curve published by \cite{Pastorello2019review} is also shown with green squares.
}
\label{fig:ngc4490_LC}
\end{figure}

In contrast with the optical follow-up, AT 2011kp remained visible in the MIR at least for eight years, until the retirement of the SST spacecraft (see the MIR light curve in Figure \ref{fig:ngc4490_LC}).
In the first observation reported by \cite{Smith2016} (at +1 year), the MIR colour is still relatively blue, at $[3.6]-[4.5]=-0.37$ mag. From 2014, the MIR light curves decline quite slowly for about four years, with a rate of $\sim0.45\pm0.04$ mag yr$^{-1}$ in both filters. In the same period, the $[3.6]-[4.5]$ colour slowly increases from 0 mag at +3 years to 0.2 mag at +6 years (corrected for an extinction of $E(B-V)=0.32$ mag, as suggested by \citealt{Smith2016} and \citealt{Pastorello2019review}). 
The trend of diminishing in brightness while becoming only mildly redder can be an indication that the emitting material is not actually cooling. This can be explained with an IR echo \citep{Dwek1983ApJ...274..175D, Graham1986MNRAS.221..789G, Meikle2006ApJ...649..332M, Mattila2008MNRAS.389..141M} from distant, pre-existing circumstellar dust surrounding the central engine of the transient, as observed in V838 Mon \citep{Crause2005MNRAS.358.1352C, Tylenda2005_V838}.
From 2018, the decline rate of the $Ch1$ light curve becomes steeper, and the $[3.6]-[4.5]$ colour also increases more rapidly to 1.1 mag at +7 years, which is remarkably similar to the colour of AT 1997bs at the same phase, while the $Ch2$ light curve does not change its slope. The $[3.6]-[4.5]$ colour curve is shown in the Appendix (Figure \ref{fig:MIR_color}).
Finally, at +7.5 years, AT~2011kp is no longer detectable in the $Ch1$ images, but it is still visible in the $Ch2$ filter, at least until +8.1 years, with the source being more than 3 magnitudes fainter than in 2012. This is the last available epoch, as SST ended its operations.

\begin{figure*}
\includegraphics[width=2.04\columnwidth]{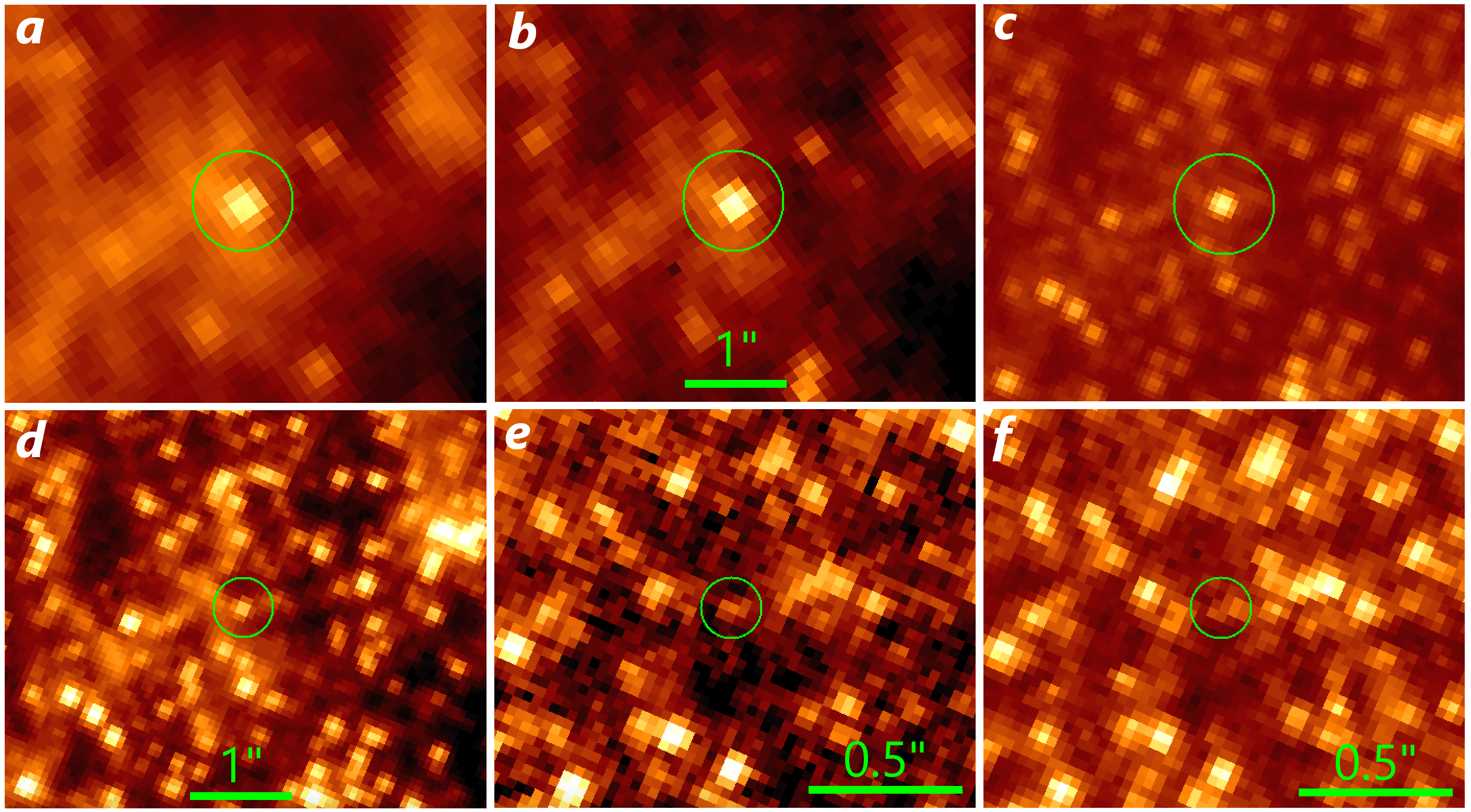}
\caption{JWST MIRI and NIRCam images of AT 2011kp. Panels are as follow: ($a$): $F770W$, ($b$): $F560W$, ($c$): $F444W$, ($d$): $F335M$, ($e$): $F187N$, and ($f$): $F115W$. In the MIR, a bright source is clearly visible 12.5 years after the first peak.
In the top row stamps, the green circles are centred at the position of the transient and have a radius of $0.5''$. In the $F335M$ stamp, the circle has a radius of $0.3''$, while in the $F187N$ and $F115W$ stamps it has a radius of $0.1''$. The latter two stamps are more zoomed-in to better identify the counterpart.
}
\label{fig:ngc4490_jwst}
\end{figure*}

The host galaxy NGC 4490 was observed by JWST/MIRI in December 2023.
AT 2011kp is well detected in both the $F560W$ and $F770W$ filters images (Fig.~\ref{fig:ngc4490_jwst}, panels $a$ and $b$), despite the observations being conducted more than 12 years after the LRN outburst.
Finally, JWST observed NGC~4490 again with NIRCam in February 2024 (12.5 years after the LRN outbursts) in multiple NIR filters from $F115W$ to $F444W$ (Fig. \ref{fig:ngc4490_jwst}, panels $c$ and $d$).
AT 2011kp is still well detectable in all filters (though it is faint in the bluest ones), and it is brighter at the longest wavelengths, indicating a red colour and a cool continuum temperature.
The $F444W$ flux aligns reasonably well with the linear decline shown by the $Ch2$ filter between 2012 and 2019 (Fig.~\ref{fig:ngc4490_LC}).
In contrast, AT~2011kp is not detected in the bluer $F200W$ and $F150W$ wide filters.
The non-detection of the remnant of AT~2011kp in the NIR, when instead it is bright in the MIR, evidences how its SED is dominated by a very cool emitting source.

\subsection{AT 2013lw (UGC 12307-OT2013)}
\begin{figure}
\includegraphics[width=1\columnwidth]{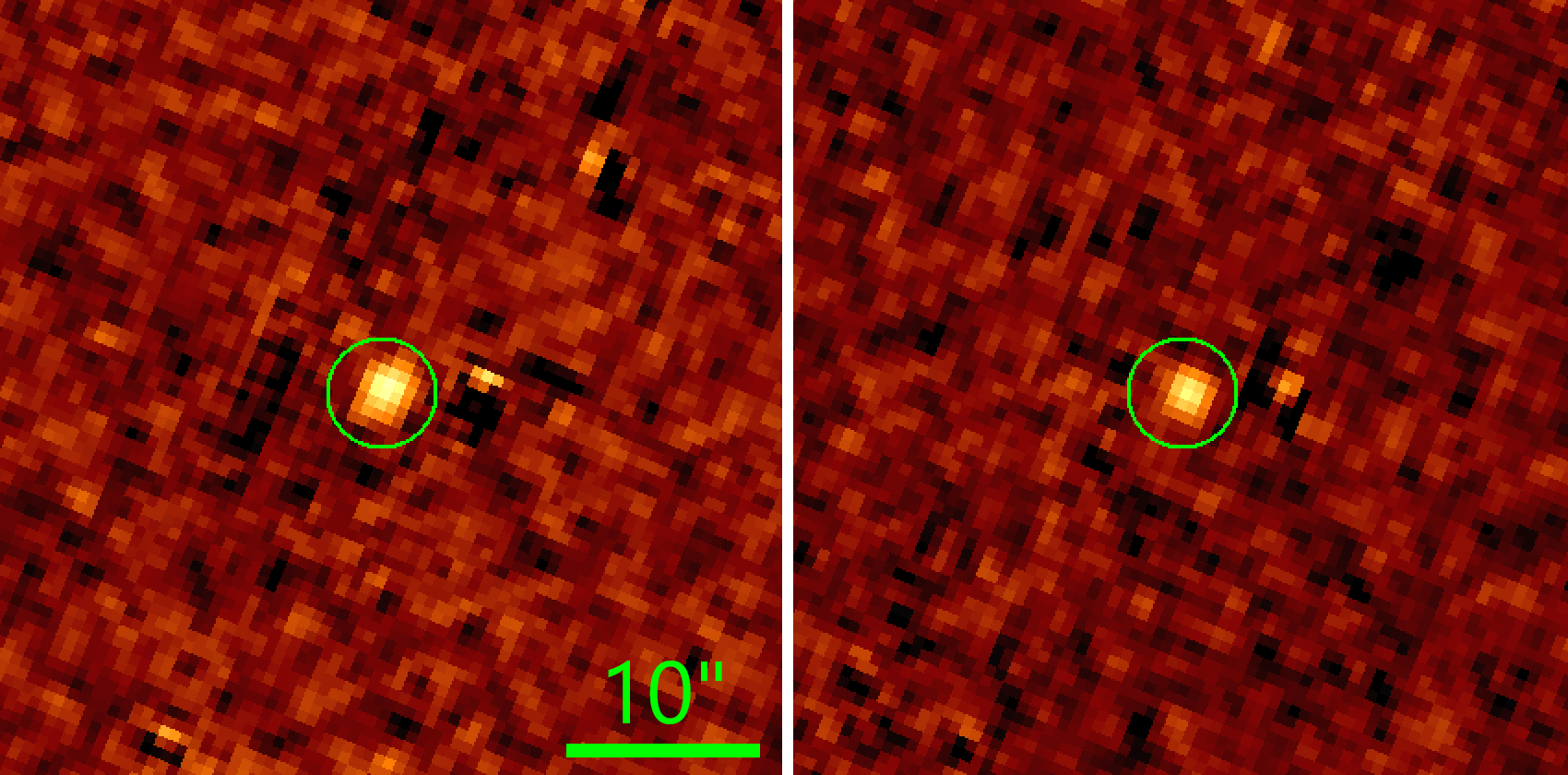}
\caption{Detection of AT 2013lw in the template-subtracted 2014 images obtained by SST in the $Ch1$ (left) and $Ch2$ (right) filters. The green circles are centred at the transient position and have a radius of $3''$.}
\label{fig:13lw_Spitzer}
\end{figure}

AT 2013lw (or UGC 12307-OT2013) was classified as a LRN by \cite{Pastorello2019review}. The epoch of the first blue peak is not known, as the object was discovered at a late evolutionary stage.
We detect AT 2013lw in SST $Ch1$ and $Ch2$ images acquired on August 2014 (13 months after the red peak), after subtracting a template SST image obtained in 2009 (Fig. \ref{fig:13lw_Spitzer}).

\subsection{AT 2015fx (\PSN)}
AT 2015fx (also known as \PSN\,and NGC 2748-2015OT1) was classified as an SN impostor by \cite{Tartaglia2015ATel.7051....1T, Tartaglia2016ApJ...823L..23T} on the basis of its peak absolute magnitude $M_r=-13.6$ and the spectral appearance. However, similarities with AT 2007sv, AT 1997bs, and AT 2014ib were noted. In fact, the light curve presented an early blue peak followed by a plateau lasting $\approx$150 days. 
The early spectra were characterised by a blue continuum and narrow Balmer lines with P-Cygni profiles, indicating low expansion velocities of $<1000$ \kms. As the spectra evolved in the following three months, the continuum become redder, while Balmer lines weakened, and absorption lines from metals started to appear. In the final spectrum, broad absorptions from TiO molecular bands also appeared, a common feature in LRNe at late phases.
\cite{Tartaglia2016ApJ...823L..23T} found a plausible progenitor star for AT~2015fx in HST/WFPC2 $F450W$ and $F814W$ images being a white-yellow supergiant of 18-20~\Msun\footnote{The mass was estimated from a comparison of the location of the progenitor within the HRD to some evolutionary tracks.}, while the majority of SN impostors tend to originate from blue stars. They also detect the progenitor in SST/IRAC images. The progenitor's properties are similar to those of another well-studied LRN, AT 2021biy \citep{Cai2022A&A...667A...4C}.
Based on the photometric and the spectroscopic similarities of AT 2015fx with other LRNe, the colour and mass matches of the progenitor with that of AT 2021biy, and its disappearance in the 2019 optical image, we propose its reclassification as a LRN.

\begin{figure}\centering
\includegraphics[width=.75\columnwidth]{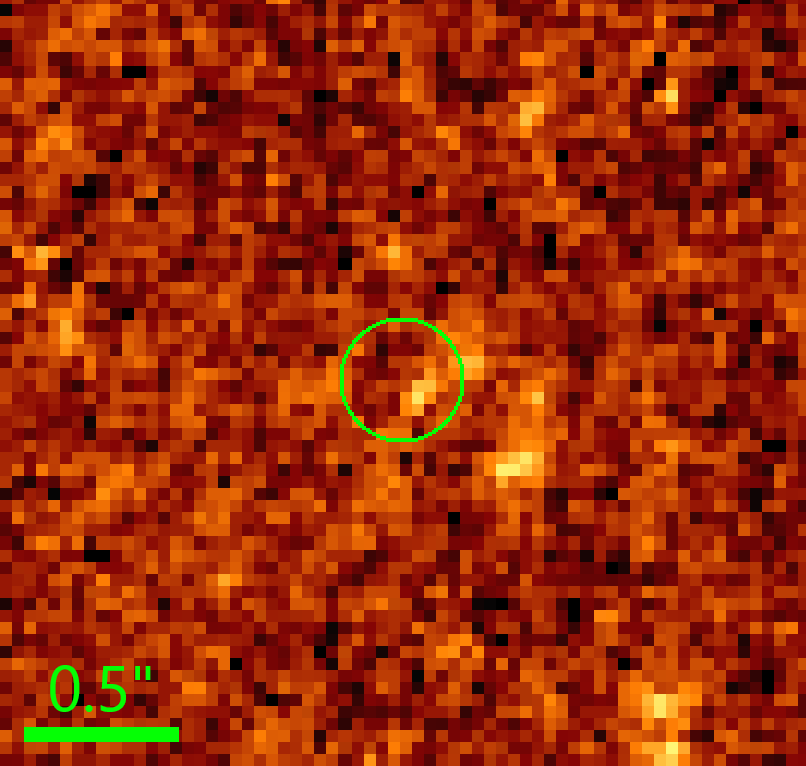}
\caption{Possible detection of AT 2015fx by HST/WFC3 in the $F814W$ filter in 2016. The green circle is centred at the transient position and has a radius of $0.2''$.}
\label{fig:PSN}
\end{figure}

We retrieved HST/WFC3 images of the field of AT 2015fx in the $F555W$ and $F814W$ filters acquired in October 2016 (1.7~years after the outburst). The transient is not detected in the $F555W$ filter. Instead, it is tentatively detected in the $F814W$-band image (Fig. \ref{fig:PSN}). This source is $\sim$2.7 mag fainter than the progenitor identified by \cite{Tartaglia2016ApJ...823L..23T} in the same filter ($F814W=24.00\pm0.18$ mag).
Recently, \cite{Baer-Way2024ApJ...964..172B} verified the disappearance of the progenitor of AT 2015fx also in the $F814W$ filter, based on a later HST/WFC3 image acquired in February 2019, although they do not report the limit, which we measured.

MIR observations of the location of AT 2015fx were performed by WISE, that targeted the field four times in 2015 and 2016. However, we do not detect the transient in the $W1$ and $W2$ filters at any epoch. The non-detection in the MIR is probably a consequence of its larger distance (24 Mpc), hence a shallower absolute magnitude detection limit. 

\subsection{AT 2015dl (M101-2015OT1)}
AT 2015dl\footnote{The object is also reported in the literature as M101-2015OT1 and M101 OT2015-1 \citep[e.g.,][]{Goranskij2016AstBu..71...82G}.} has been extensively studied by \cite{Blagorodnova2017}. They found its progenitor to be an F-type yellow supergiant with an estimated mass of 18 \Msun, which disappeared in the optical bands at late times, as proved by their deep images from the Keck telescope. To improve the characterisation of this LRN, in this paper we analyse more recent ground-based optical and NIR images, along with a much wider dataset in the MIR domain. In particular, we update the light curve of AT~2015dl in the MIR bands after a re-analysis of the SST data published by \cite{Blagorodnova2017}, and extend the photometric monitoring by further four years, including previously unpublished WISE data and more recent SST observations.

We retrieved level-2 calibrated $r$-band images of the field of AT 2015dl obtained with the 3.58m Canada France Hawaii Telescope (CFHT) with MegaPrime. These images, acquired on 2017 March 22 and 23 (+2 years), are deep and taken with excellent seeing conditions ($\sim 0.45''$). We calibrate them against the Pan-STARRS1 catalogue \citep{Chambers2016arXiv161205560C}.
We do not detect any source down to $r>25.5$ mag, which is a much fainter limit than the quiescent progenitor level reported by \cite{Blagorodnova2017} ($\sim21$ mag). 

We also retrieved a single $K$-band image from the UKIRT telescope with the WFCAM instrument obtained on 2019 June 14 (+4.3 years), which was calibrated against the 2MASS catalogue \citep{Cutri2003tmc..book.....C}. No source is detected to a limit of $K>20.8$ mag (AB scale). While this limit is brighter than the quiescent progenitor ($K=21.44\pm0.11$, converted to AB, \citealt{Blagorodnova2017}), it confirms the luminosity decline of the transient in the NIR region a few years after the outburst.

\begin{figure}
\includegraphics[width=1\columnwidth]{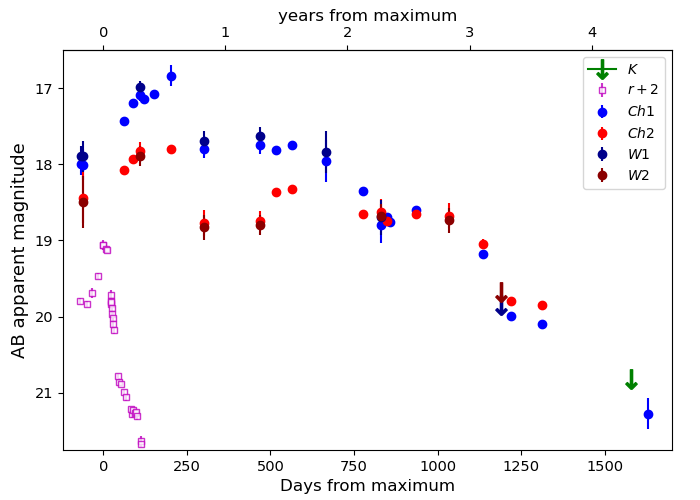}
\caption{Mid-infrared SST and WISE light curves of AT 2015dl spanning over 4.5 years of evolution.
A $K$-band upper limit from UKIRT/WFCAM is also shown.
Downwards arrows mark the upper limits.
For reference, the $r$-band light curve published by \cite{Blagorodnova2017} is also shown with magenta squares.
}
\label{fig:15dl_LC}
\end{figure}

SST observed the region multiple times between 2015 and 2019. Although the data obtained from April to September 2015 were already published by \citealt{Blagorodnova2017}, for consistency, we remeasured them. Our measurements well matches theirs. AT 2015dl remains visible in the MIR for years after the outburst, as is still detected nearly five years after the first maximum. The final MIR light curve is shown in Figure \ref{fig:15dl_LC}. 
The field was also imaged twice by SST on 2004 March 8 - during the Cold Mission - with all IRAC filters, hence the two exposures in the same filters could be combined to obtain deeper upper limits to the progenitor luminosity.

WISE monitored the field around AT 2015dl in multiple occasions between December 2014 (i.e. 2 months before the optical peak) and March 2018 (3 years later). Given the modest space resolution of the WISE observation, the photometric data were obtained after the subtraction of the host galaxy contamination. As a template, we used WISE images obtained on June 2010.
The first observation of WISE, on 2014 December 19, was obtained about 50 days before the official announcement of the LRN discovery (on 2015 February 10), hence during the slow luminosity rise before the LRN outburst onset. The $W1-W2$ colour during the pre-outburst rise is $\sim-0.6$ mag.

\begin{figure}\centering
\includegraphics[width=.75\columnwidth]{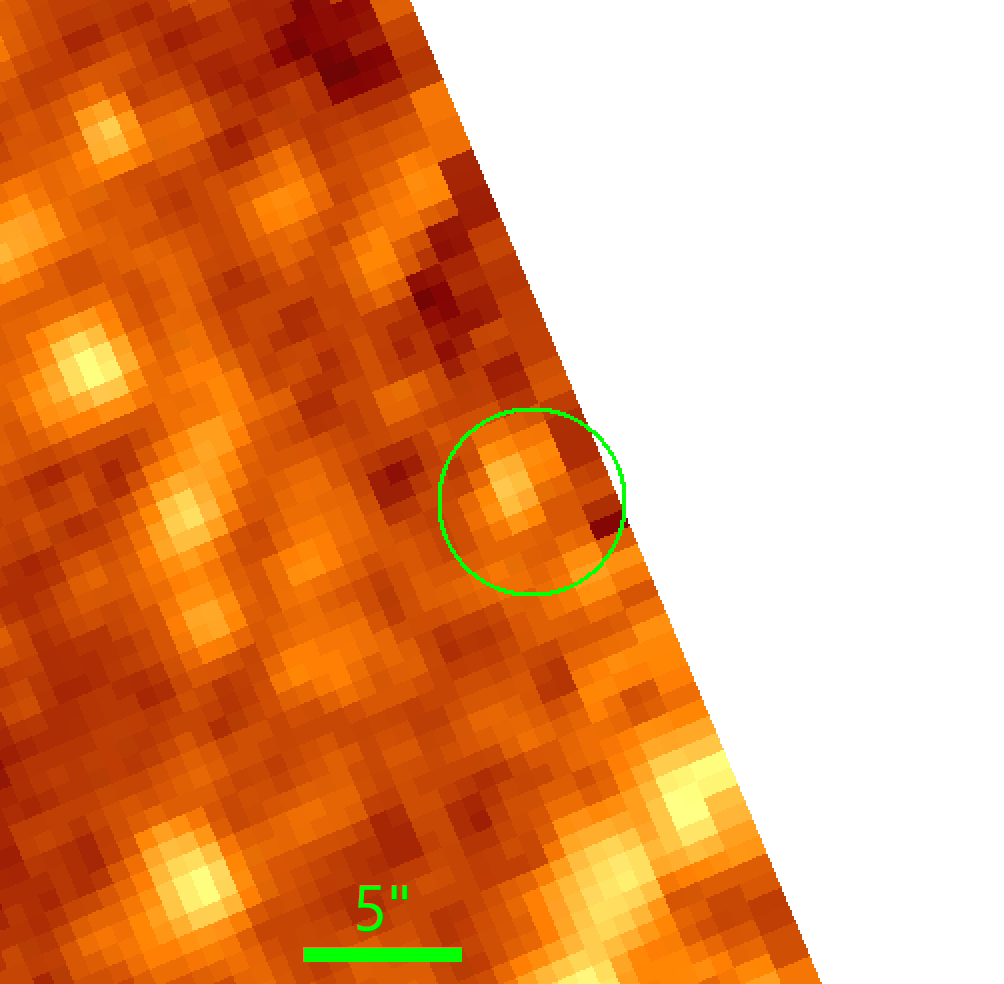}
\caption{AT 2015dl in the SST image obtained on 2019 August 2. Despite being imaged at the edge of the frame, the source is still detectable. The green circle, centred at the transient position, has a radius of $3''$.
}
\label{fig:15dl_Spitzer}
\end{figure}

In 2015, AT 2015dl is observed to brighten in the MIR, reaching a maximum about eight months after the $r$-band peak. Later, the light curve starts a general declining trend, with some fluctuations (Fig. \ref{fig:15dl_LC}).
Another SST epoch with both $Ch1$ and $Ch2$ filters was acquired on 2018 September 18, and a final $Ch1$ observation was obtained on 2019 August 2. In the latter image, while the source is located at the edge of the frame (see Fig.~\ref{fig:15dl_Spitzer}), it is $\sim$4 mag fainter than at maximum, and just 0.5 mag brighter than the upper limit on the progenitor obtained in the same filter from the 2004 SST observations.

At the time of the $r$-band observation (+2 years), the $r-[3.6]$ colour of AT 2015dl is at least 7.2 mag. Very high optical--IR colours are usually interpreted as an indication of dust condensation. While the object is not detected in the optical, it is still bright in the SST $Ch1$ filter.
This is similar to the behaviour shown by V1309 Sco in the optical/infrared bands \citep{Nicholls2013MNRAS.431L..33N, Tylenda2016A&A...592A.134T}. 

Starting from +2 years, we note a dramatic evolution in the MIR colour: The $[3.6]-[4.5]$ colour changes from $\sim -1.0$ mag (this colour index remains stable between +6 months and +1.5 years), to $\sim-0.3$ mag at +775 days, and then rises to $+0.0$ mag at +850 d. Finally, the colour continues to increase up to $[3.6]-[4.5]= +0.3$ mag in the last available observation with the $Ch2$ filter, at +3.6 years. Together with the continuous decline in luminosity observed in both filters, this large colour variation is an indication of a cooling in the expanding material. The MIR $[3.6]-[4.5]$ and $W1-W2$ colour curves are shown in the Appendix (Fig. \ref{fig:MIR_color}). 

\subsection{AT 2018bwo}
This LRN was studied in detail by \cite{Blagorodnova2021} and \cite{Pastorello2023}. Its progenitor was identified as a yellow giant of $M\approx$11-16 \Msun\,by \cite{Blagorodnova2021}.

We retrieved an HST/WFC3 image acquired with the $F350LP$ filter on October 2022. The filter is very wide and covers the entire optical range\footnote{https://hst-docs.stsci.edu/wfc3ihb/appendix-a-wfc3-filter-throughputs/a-2-throughputs-and-signal-to-noise-ratio-data/uvis-f350lp}, from 3500 to 10000 \AA, hence is roughly equivalent to a Clear filter image. A source is detected at the position of the candidate progenitor detected by \cite{Blagorodnova2021} in HST images acquired in 2004.
Although the source is much fainter than the progenitor (the optical magnitude was $\sim23$) and no colour information is available, this late detection (at an absolute magnitude of $M_{F350LP}\sim-1.7$ mag) indicates that four years after the outburst the source at the location of AT~2018bwo is fainter than the quiescent progenitor, at least in the optical domain.

\begin{figure}\centering
\includegraphics[width=.75\columnwidth]{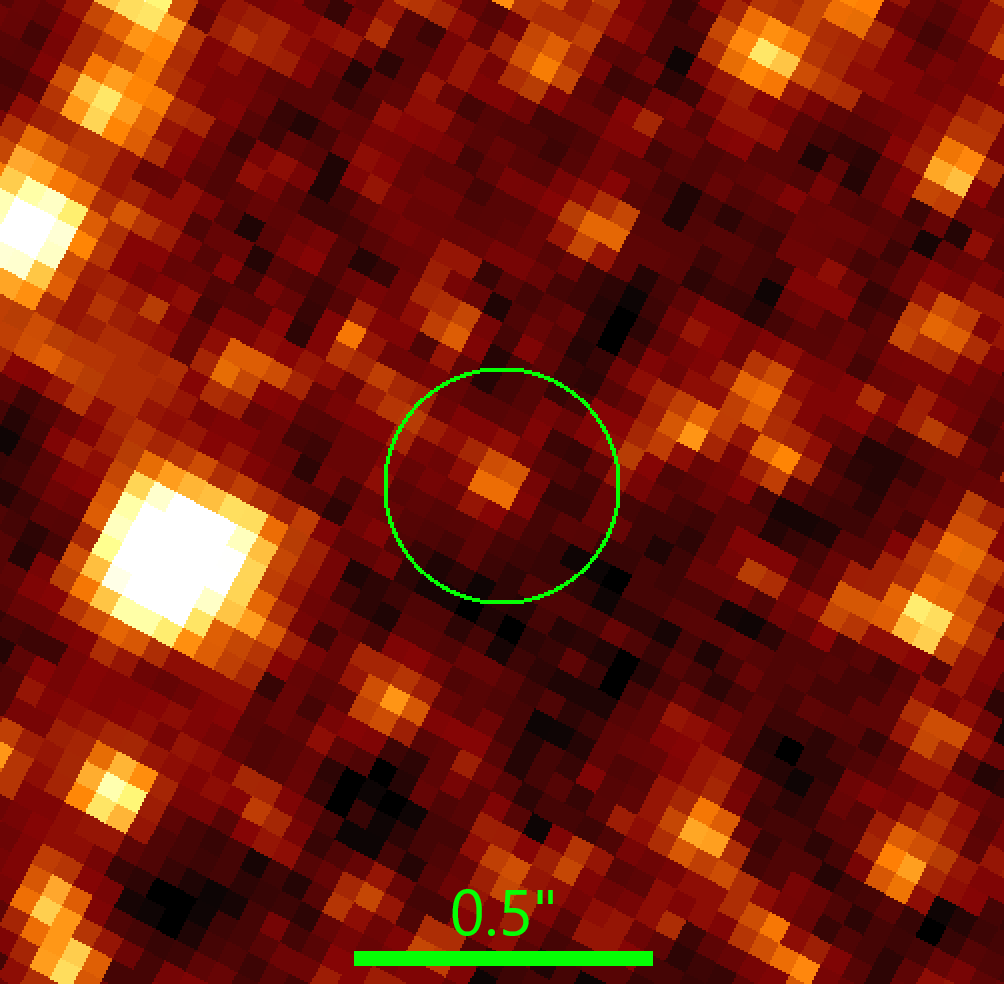}
\caption{HST/WFC3 image of the field of AT 2018bwo taken with the $F350LP$ filter on 2022 October 1. 
The green circle is centred at the transient position and has a radius of $0.2''$.}
\label{fig:18bwo_HST}
\end{figure}

\subsection{AT 2019zhd}
AT 2019zhd is the third LRN observed in the Andromeda Galaxy, after M31-RV \citep{Mould1990ApJ...353L..35M, Boschi2004AA...418..869B} and M31-LRN2015, \citep{Kurtenkov2015A&A...578L..10K, Williams2015ApJ...805L..18W, MacLeod2017ApJ...835..282M, Lipunov2017MNRAS.470.2339L, Blagorodnova2020}.

\begin{figure}
\includegraphics[width=1\columnwidth]{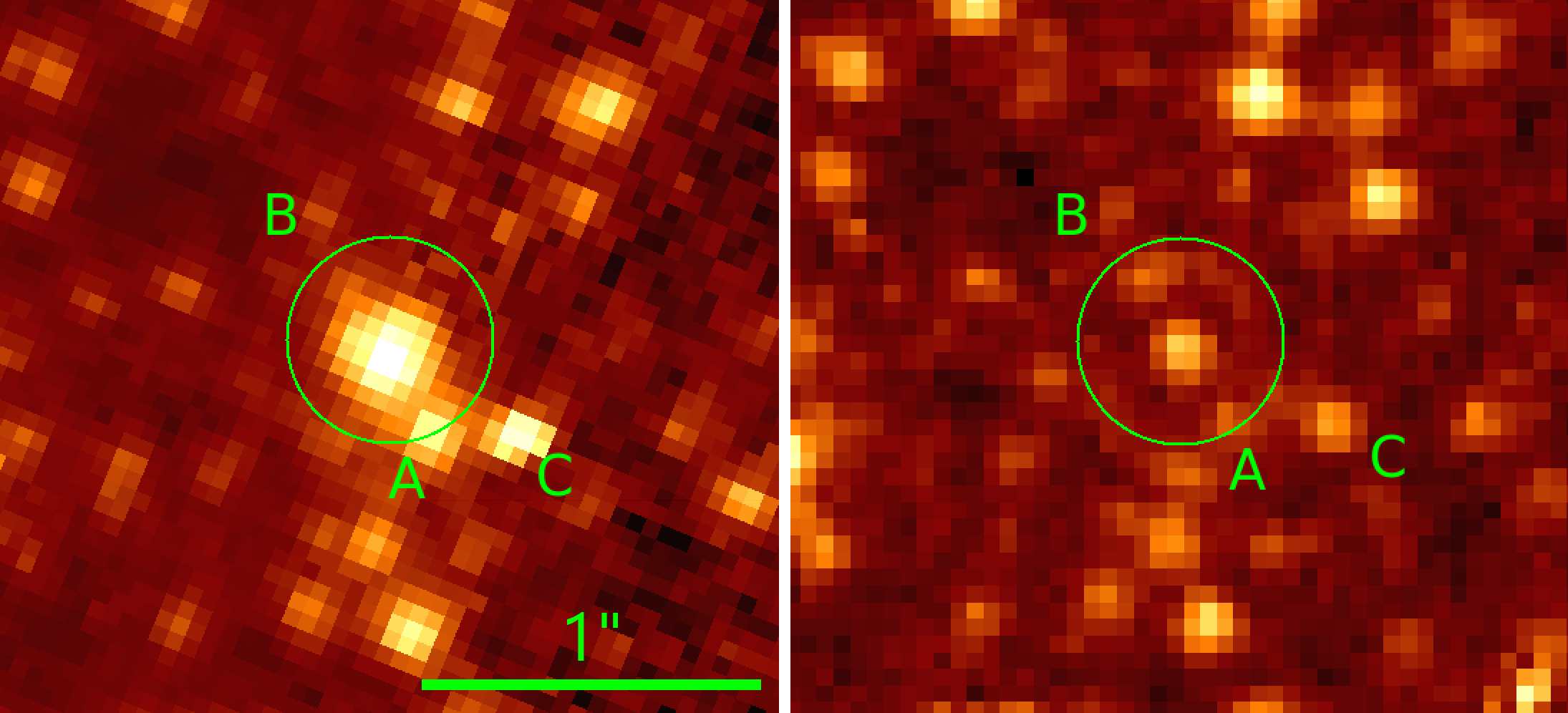}
\caption{HST/ACS images taken in 2022 of the field of AT~2019zhd. Left: Detection in the $F814W$ filter. Right: Detection in the $F475W$ filter. The green circles are centred at the transient position and have a radius of $0.31''$.
The three sources identified by \cite{Pastorello2021_19zhd} are also labelled. Notably, none of them match the position of AT 2019zhd.}
\label{fig:19zhd_HST}
\end{figure}

\cite{Pastorello2021_19zhd} conducted a detailed study on this object, and in particular analysed a set of HST/WFPC2 images acquired in 1997 searching for the progenitor stellar system. These images did not allow the authors to securely identify a progenitor star. At the exact LRN coordinates, no source was visible down to limiting magnitudes of $F439W\sim$ 25.6 mag, $F814W\sim$ 25.7 mag, and $F814W\sim$ 25.4 mag.
However, three sources were found in the proximity of the LRN location, labelled as sources A, B and C in that paper. None of their position perfectly matched that of the LRN centroid \citep[see Fig. 9 in][]{Pastorello2021_19zhd}.
However, among the three sources, only source C was outside the $0.31''$ error-box radius. For this reason, the other two sources were still considered potential progenitor candidates, with source A being the most plausible. 
We retrieved HST/ACS images acquired in January 2022 (+1.9 yrs) of the field of AT 2019zhd, with the $F814W$ filter. The transient is clearly detected (Figure \ref{fig:19zhd_HST}, left panel), within the $0.31''$ positional uncertainty mentioned above, and is clearly distinct from the three sources identified by \citet{Pastorello2021_19zhd}. For this reason, we can now safely rule out any of them being the progenitor of LRN AT~2019zhd.

AT 2019zhd is also detected by HST/ACS in the $F475W$ filter in July 2022. Although the detection of a source in a blue filter such as $F475W$ is somewhat unusual if it is obscured by dust, the $F475W-F814W$ colour (close to Sloan $g-i$) estimated for AT 2019zhd in this phase is $4.4\pm0.1$ mag (corrected for an extinction of $E(B-V)=0.055$ mag).
At the distance of M~31 (0.78 Mpc), the absolute magnitudes of AT 2019zhd in 2022 are $M_{\rm F814W}=-3.94$ and $M_{\rm F475W}=+0.42$ mag. These parameters of colour and luminosity are consistent with those of a M6-9 giant star as the survivor of the LRN event, as we will determine in Sect. \ref{sect:SEDs}.

While AT 2019zhd is not detected by WISE on 2019 December 26, at a time close to the first blue peak (template from January 2014 with limits: $W1>18.8$, $W2>18.3$ mag), it is seen in the $W1$ filter in July and December 2019, 7 and 12 months after the outburst, respectively.
This late detection in the MIR is remarkable considering that the optical follow-up of the transient lasted less than two months.

\subsection{AT 2020hat}
AT 2020hat was studied by \cite{Pastorello2021_20hat}, and in 2017 HST/ACS images they found a K-supergiant star as a possible progenitor.
We retrieved SST images acquired between 2007 and 2019 in $Ch1$ and $Ch2$. In the 2007 images, we estimate profound limits to the progenitor brightness. We use these images as templates for more recent observations. 

\begin{figure}
\includegraphics[width=1\columnwidth]{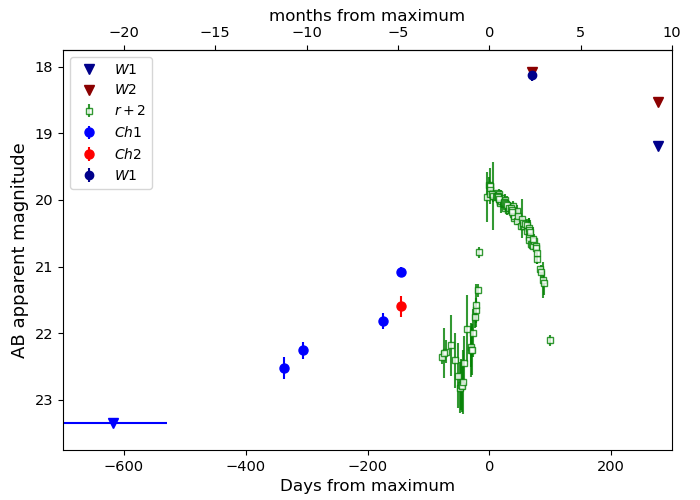}
\caption{Mid-infrared light curve of AT 2020hat before (from SST) and during (from WISE) the outburst. Downwards triangles mark upper limits. For reference, the $r$-band light curve from \cite{Pastorello2021_20hat} is also shown with green squares.}
\label{fig:20hat_LC}
\end{figure}

\begin{figure}
\includegraphics[width=1\columnwidth]{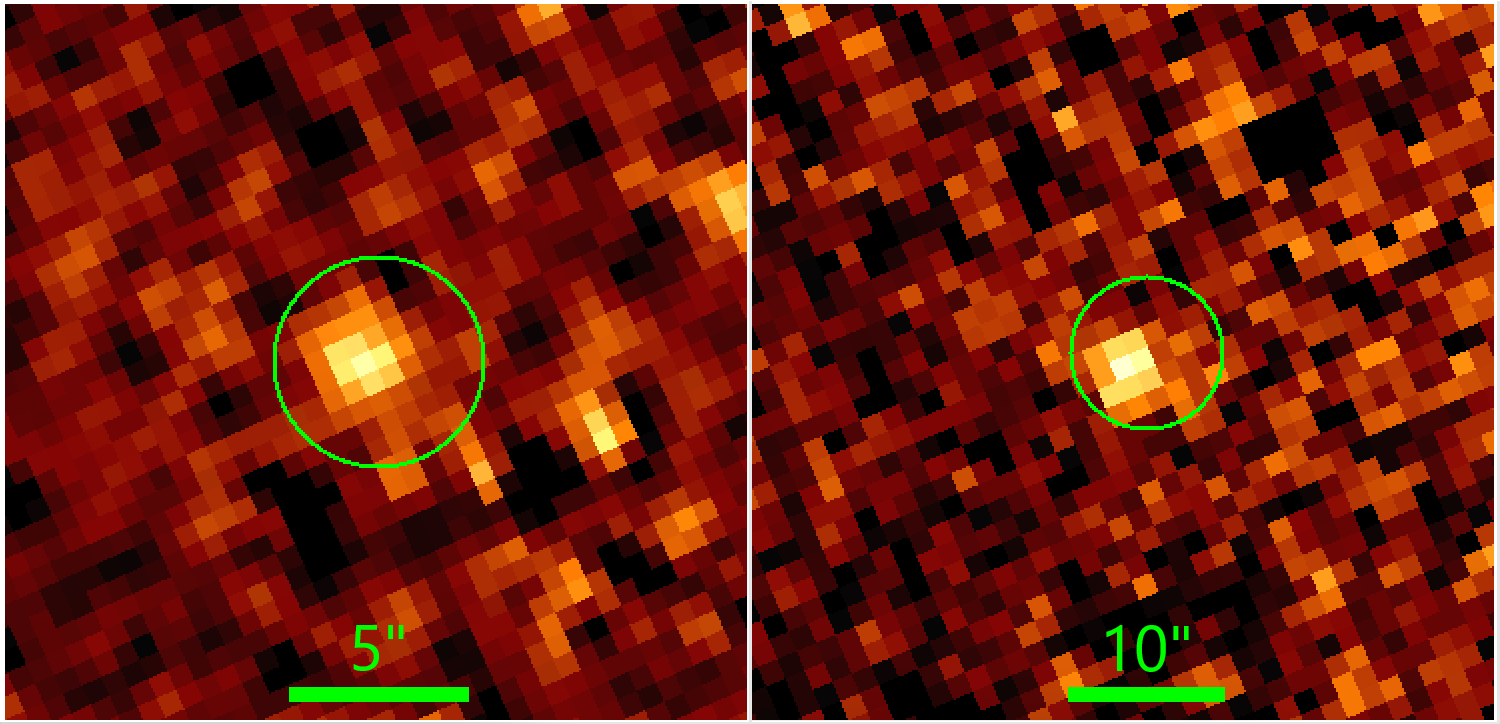}
\caption{Detection of AT 2020hat in the MIR template-subtracted images. Left: Last $Ch1$ observation from SST (on 2019 November 21). Right: WISE $W1$ observation performed on 2020 June 24. The green circles are centred at the transient position and have radii of $3''$ and $10''$, respectively.}
\label{fig:20hat}
\end{figure}

Between 2014 and 2018, upper limits on the difference images are obtained after the subtraction of the 2007 templates. 
Later, between May and November 2019 (before the end of the mission), we observe a slow rise in luminosity in the final four SST observations with the $Ch1$ filter (Fig. \ref{fig:20hat_LC}).
This slow luminosity rise observed in the MIR mirrors the one observed in the optical filters prior to numerous LRNe. However, so far, AT~2020hat is the only extragalactic LRN for which the pre-outburst rise is observed in the infrared domain.
A single $Ch2$ image is available at the same epoch as the last $Ch1$ observation (Fig. \ref{fig:20hat}, left panel). From this point, the $[3.6]-[4.5]$ colour is $\sim -0.52 $ mag, hence the object was remarkably blue at that phase, similarly to AT 2015dl. This is an indication that the pre-outburst IR emission likely originated from a hot photosphere rather than from cold, pre-existing dust.

Two epochs of WISE imaging were obtained after the outburst (templates from January 2014). 
In the first epoch (2020 June 24, hence 70 d post $r$-band maximum), we detect a source at the LRN position only in the $W1$ filter (Fig. \ref{fig:20hat}, right panel), 4.3 magnitudes brighter than the earliest $Ch1$ measurement (obtained 11 months prior). This is similar to the steep luminosity jump observed in the optical bands from the pre-outburst level to the maximum light \citep{Pastorello2021_20hat}.
Instead, at the same epoch, an upper limit is inferred in the $W2$ filter, which indicates the source was rather blue around peak, with a colour of $(W1-W2)_0<0.05$ mag.
Finally, nine months after the $r$-band maximum (in January 2021), a second epoch of WISE imaging was obtained, and only upper limits are measured in both filters, indicating the source has faded by at least 1.1 mag also in $W1$ seven months after the previous measurement. The full MIR light curve is shown in Figure \ref{fig:20hat_LC}.

The host galaxy NGC 5068 was imaged by JWST in 2023 in multiple filters. Unfortunately, the location of AT 2020hat was just outside the images footprint, and thus the object was not observed.

\begin{figure}\centering
\includegraphics[width=\columnwidth]{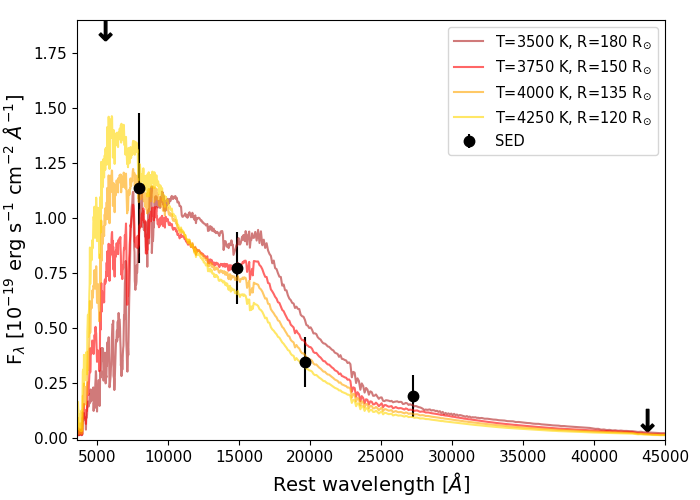}
\includegraphics[width=.78\columnwidth]{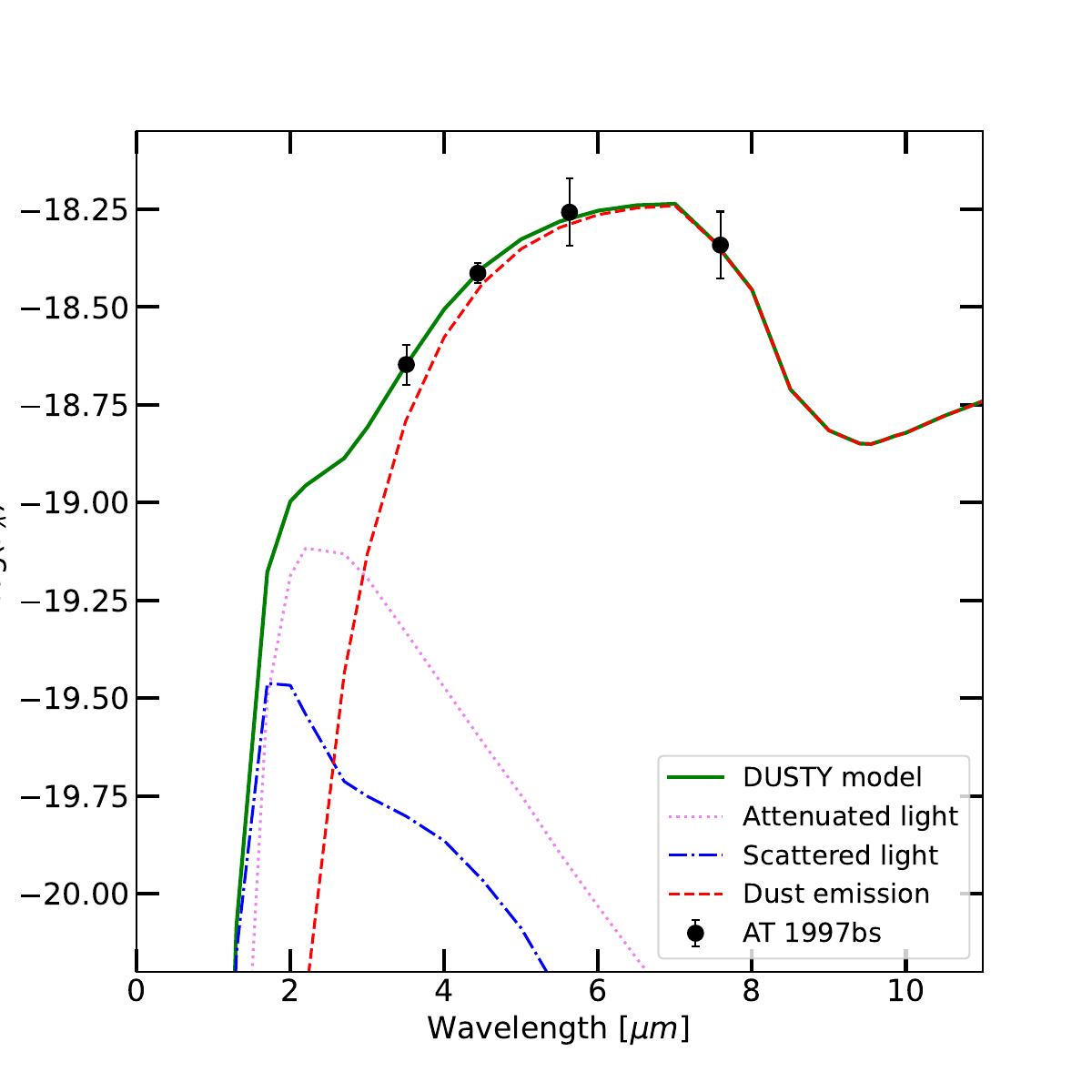}
\includegraphics[width=.78\columnwidth]{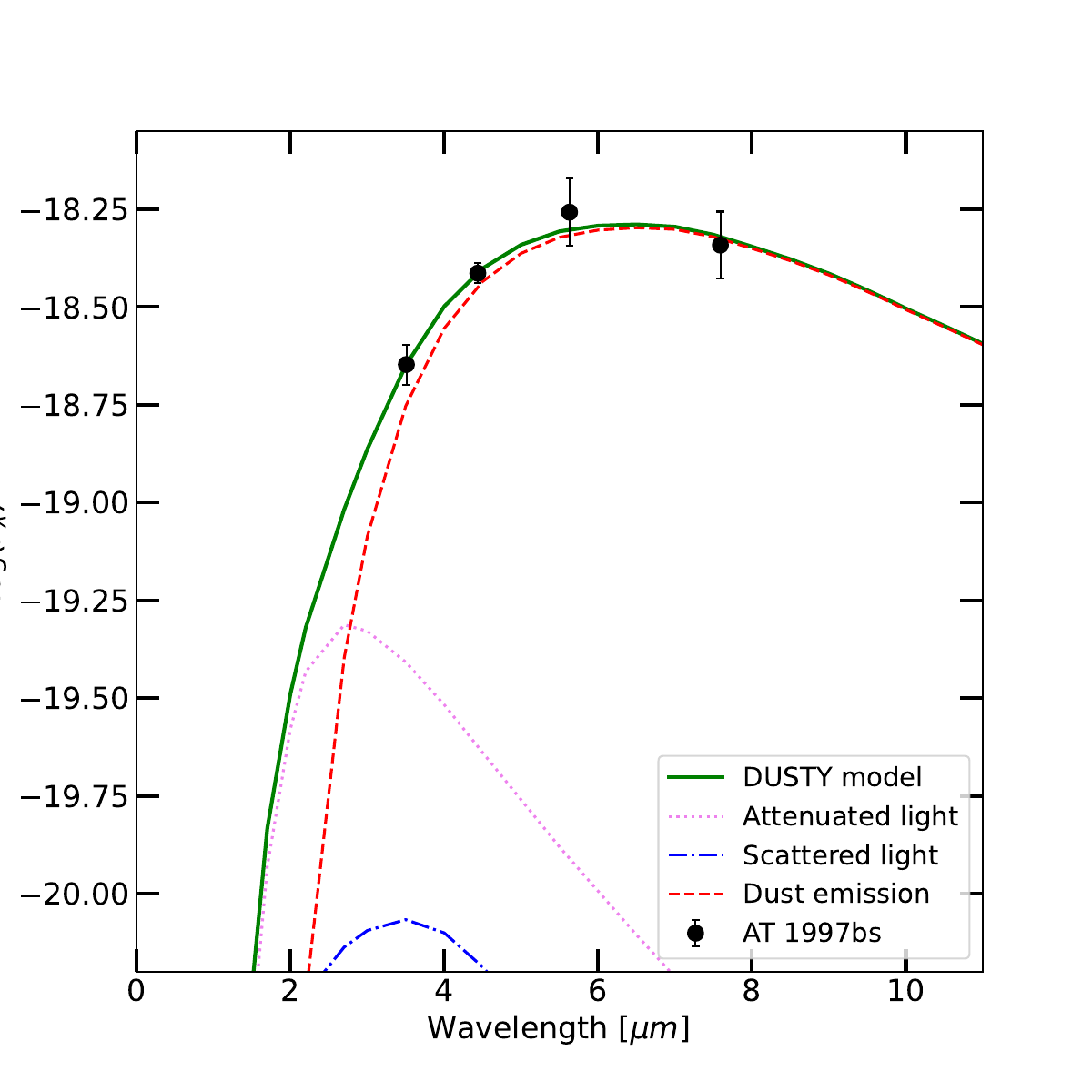}
\caption{SEDs of AT 1997bs. 
Top panel: SED $\sim27$ years after the outburst from HST/ACS and JWST/NIRCam observations. The SED is compared to ATLAS9 stellar atmospheric models from \cite{Castelli2003IAUS..210P.A20C}. Good matches are found with models with temperatures between 3750 and 4250 K, and radii between 150 and 120 \Rsun.
Middle panel: SED at the time of the 2004 SST observations. This cooler SED is fitted with a DUSTY model with a 100\% silicate composition, a dust temperature of $T_d=600$ K, a source temperature of $T_s=4000$ K, and an optical depth of $\tau_V=68.5$.
Bottom panel: Same but for a model with a 100\% AMC composition and optical depth of $\tau_V=16$.
}
\label{fig:SED_97bs}
\end{figure}

\section{Discussion}\label{sect:discussion}

The majority of the LRNe analysed in this work had relatively massive progenitors, in the range 10 to 25 \Msun, as inferred from the publications on individual objects. 
The only exception is AT~2019zhd which, given its low peak luminosity of $-9.6$ mag, is most likely a low mass object (3.4 \Msun\,as reported by \citealt{Pastorello2023}; 6 \Msun\,according to \citealt{Chen2024ApJ...963L..35C}). Nonetheless, its fate seems to be similar to that of LRNe produced by more massive systems (see below).
These masses were determined through archival images from HST by measuring the luminosity and colour of the progenitor system, positioning them in an HRD and compare with evolutionary tracks of binaries. Another approach is to implement the empirical relation between $V$-band absolute magnitude at the second peak and progenitor mass revised by \cite{Cai2022A&A...667A...4C}.

A general trend that seems to emerge from this research is that LRNe tend to fade well below the progenitor level in the optical bands, while remaining bright in the NIR and (even more) in the MIR for many years after the outburst.
An high IR luminosity seems to be an ubiquitous feature of the light curves of LRNe at late phases, determining a shift of the SED peak towards the MIR domain.
The very late time detections of LRNe in the IR domain can be easily explained with the formation of a conspicuous amount of new dust in the months/years after the outburst.
This statement is also supported by the evidence of a blue-shift of the spectral lines observed in the late-time spectra of LRNe \citep[e.g.,][]{Mason2022A&A...664A..12M}.
For instance, M31-LRN2015 (not included in our sample), was detectable in the $K$-band for more than two years, and was still detected by \cite{Blagorodnova2020} in the SST $Ch2$ filter nearly five years after the outburst, while it was below the detection threshold in the optical and NIR bands, as well as in the SST $Ch1$ images.

\subsection{Spectral energy distributions of the remnants}\label{sect:SEDs}
\begin{figure}\centering
\includegraphics[width=\columnwidth]{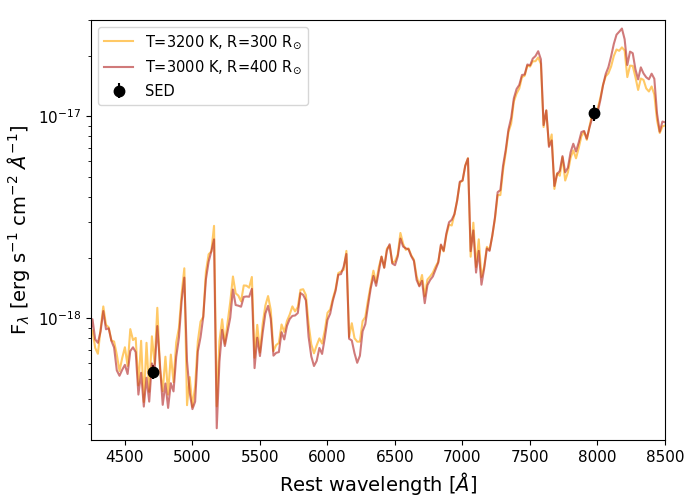}
\caption{Spectral energy distribution of the remnant of AT~2019zhd compared to PHOENIX stellar atmospheric models from \cite{Husser2013A&A...553A...6H}. The PHOENIX spectra have been convolved with a Gaussian kernel with a width of 20 \AA. Matches are found with models with temperatures of 3000 and 3200 K, and radii of 400 and 300~\Rsun.}
\label{fig:SED_AT2019zhd_PHOENIX}
\end{figure}

We constructed the SED of AT 1997bs in 2023-2024 by combining the information from the HST ($F814W$) and JWST filters ($F150W$, $F200W$, $F277W$, plus the limit in $F444W$). 
We compared it with the ATLAS9 stellar atmospheric models by \cite{Castelli2003IAUS..210P.A20C}\footnote{retrieved here: \url{https://archive.stsci.edu/hlsps/reference-atlases/cdbs/grid/ck04models/}}. We adopted the models with $Z/Z_{\odot}=1$ (hence log$(Z) =+0.0$) and log$(g) =1.0$. 
We find good matches for models with photospheric temperatures between 3750 and 4250 K, and stellar radii between 150 and 120 \Rsun\,(for the hottest and coldest models, respectively)\footnote{We warn that these parameters are obtained after correcting for a reddening of only $E(B-V)=0.04$ mag. We cannot exclude, given the earlier observations, that a significant amount of dust is still present and can obscure the merger, making it redder and fainter.}. Their corresponding bolometric luminosities are $(1.61 \pm 0.06 \pm 0.10)\times10^{37}$ erg~s$^{-1}$ (hence, log($L/L_{\odot})=3.62 \pm 0.02 \pm 0.03$). The matches of the ATLAS9 models to the remnant's SED are shown in Fig. \ref{fig:SED_97bs}, left panel.
These parameters are consistent with those of a red giant star as the outcome of the AT 1997bs merging event, thus ruling out it was a terminal SN explosion or an SN impostor from a blue massive star, such as an LBV.
Recently, \cite{Steinmetz2025A&A...699A.316S} also constrained the outcome of the Galactic LRN OGLE-2002-BLG-360 to be a red giant star, with an effective temperature of 3200~K and a radius of 300~\Rsun.
Finally, from our +3 yrs HST observations of AT 2019zhd, we find that its remnant is also compatible with the colour and luminosity of a red giant star (unless the source is heavily extincted).
We stress that this red giant state is the outcome of the coalescence of two stars, and not the traditional stellar evolution determined by changes in nuclear core burnings.
We compared the SED of the outcome of AT 2019zhd with the PHOENIX stellar atmospheric models by \cite{Husser2013A&A...553A...6H}\footnote{retrieved here: \url{https://pollux.graal.univ-montp2.fr/collections/}}. We adopted the models with $[Fe/H]=0.0$ and log$(g) =1.0$. 
We find a good match for the models with photospheric temperature of 3000 and 3200~K, and stellar radii of 400 and 300 \Rsun, respectively. Their corresponding bolometric luminosities are log$(L/L_{\odot})=4.07$ and $3.93$, respectively. The matches of the PHOENIX models to the remnant's SED are shown in Fig.~\ref{fig:SED_AT2019zhd_PHOENIX}.
This effective temperature of $T=3100\pm100$ K means that the object, if it is a star, might be located to the right of the Hayashi line, hence not (yet) in hydrostatic equilibrium, as it was observed only three years after the merger, or it is affected by dust extinction.

\subsection{The dusty environment}
From the +27 yrs SED of AT~1997bs, we also note the lack of a MIR excess. In fact, the source is not detected at 4.4 $\mu$m, hence the dust formed years prior (responsible for the SST detection in 2004) either evaporated in the subsequent two decades, or the dust temperature has largely decreased, moving the peak of a putative cold SED component towards longer wavelengths.
For instance, JWST observations of SN 2014C after 10 yrs reveal that the dust has cooled to temperatures of 200-300 K \citep{Tinyanont2025ApJ...985..198T}, and therefore the SED would peak at 10-15 $\mu$m.

The outburst of AT 2011kp occurred in a widely observed galaxy, giving us one of the rare opportunities to obtain well-sampled IR light curves for a LRN. In particular, we can study the temporal evolution of its SED, hence constraining the evolution of the temperature of the different emitting regions, including the coldest component which is attributed to dust emission. \cite{Smith2016} analysed the SED of AT~2011kp at 900-1000 days ($\sim3$ years) after the LRN onset, estimating the temperature of the dust to be 725 K and constraining its location at a distance of 80 AU. 

We can update the dust parameters characterisation by constructing the SED of AT 2011kp using more recent SST $Ch1$ and $Ch2$ observations obtained in 2018 (7 years after the LRN outburst), when the object shows a very red MIR colour, $[3.6]-[4.5]=+1.6$ mag. Unfortunately, only two filters are available to constrain the temperature, hence the results are necessarily very uncertain. The extinction-corrected two-band SED is fitted by a cold black body (BB) (only due to the lack of more points) with an effective temperature of 508$\pm$55~K and a radius of 210$\pm$60~AU, suggesting that the dust temperature progressively declines, while the radius of the dust emitting region increases.

Adopting the same approach for all the SST epochs, we can actually reconstruct the evolution of the radius, temperature and luminosity of the BB. The corresponding curves are plotted in Fig.~\ref{fig:RTL_11kp}. In particular, the evolution of the radius mirrors that of the colour curve. The object is very hot ($T\sim5000-6000$~K) and luminous ($\sim10^{41}$ erg s$^{-1}$) at +1 year, while it cools down to $\sim1000$~K and becomes much dimmer between +2.5 and +6 years.
Looking again at Fig.~\ref{fig:RTL_11kp}, we note that $R_{BB}$ of AT~2011kp between +2 and +6 yrs is roughly constant, while $T_{BB}$ is slowly decreasing. We fitted the temperature evolution with a function $At^{-0.5}$, which provides a good fit ($\chi^2_{d.o.f.}\simeq2.63$) to the data at those phases. Both are theoretical expectations for an IR echo from pre-existent dust \citep{Smith2016}. 
Instead, between +6 and +7 yrs, $T_{BB}$ drops faster, while $R_{BB}$ is rapidly increasing, and the MIR colour also increases (Fig.~\ref{fig:MIR_color}). All together, these are observational evidences of rapid dust formation \citep{Kotak2009ApJ...704..306K, Meikle2011ApJ...732..109M, Gall2014Natur.511..326G}.

The most recent observations publicly available of the AT~2011kp site were obtained by JWST in late 2023 and early 2024, about 13 years after the LRN outburst. Merging those data obtained in multiple bands, we construct a well-sampled SED. AT 2011kp is a strong emitter in all filters at wavelengths $>3\,\mu$m, especially at 7.7$\mu$m. 
Another source is visible at only $0.08''$ (3.5 pc at the distance of NGC~4490) east of AT 2011kp in all the NIR filters (see Figure \ref{fig:ngc4490_jwst}, panel $f$).
AT~2011kp is also detected in the $F187N$ (Fig. \ref{fig:ngc4490_jwst}, panel $e$), which covers the region of the Paschen $\alpha$ emission line.
We remark that the LRN spectra at late phases, along with molecular bands of TiO and VO, show clear \Ha emission lines (see \citealt{Pastorello2019review}). This makes the contribution of the Paschen~$\alpha$ emission line plausible 
in the late detection in the $F187N$ band\footnote{We should also account that JWST observations are not affected by the strong telluric absorption bands that usually suppress the Paschen~$\alpha$ emission in ground-based telescope observations \citep[e.g.,][]{Blagorodnova2021, Pastorello2023}.}.
Finally, we note that in the $F187N$-band image the source at $0.08''$ from the LRN location is not unequivocally detected. This suggests that the source in the proximity of AT~2011kp has a stellar continuum.

\begin{figure}\centering
\includegraphics[width=.78\columnwidth]{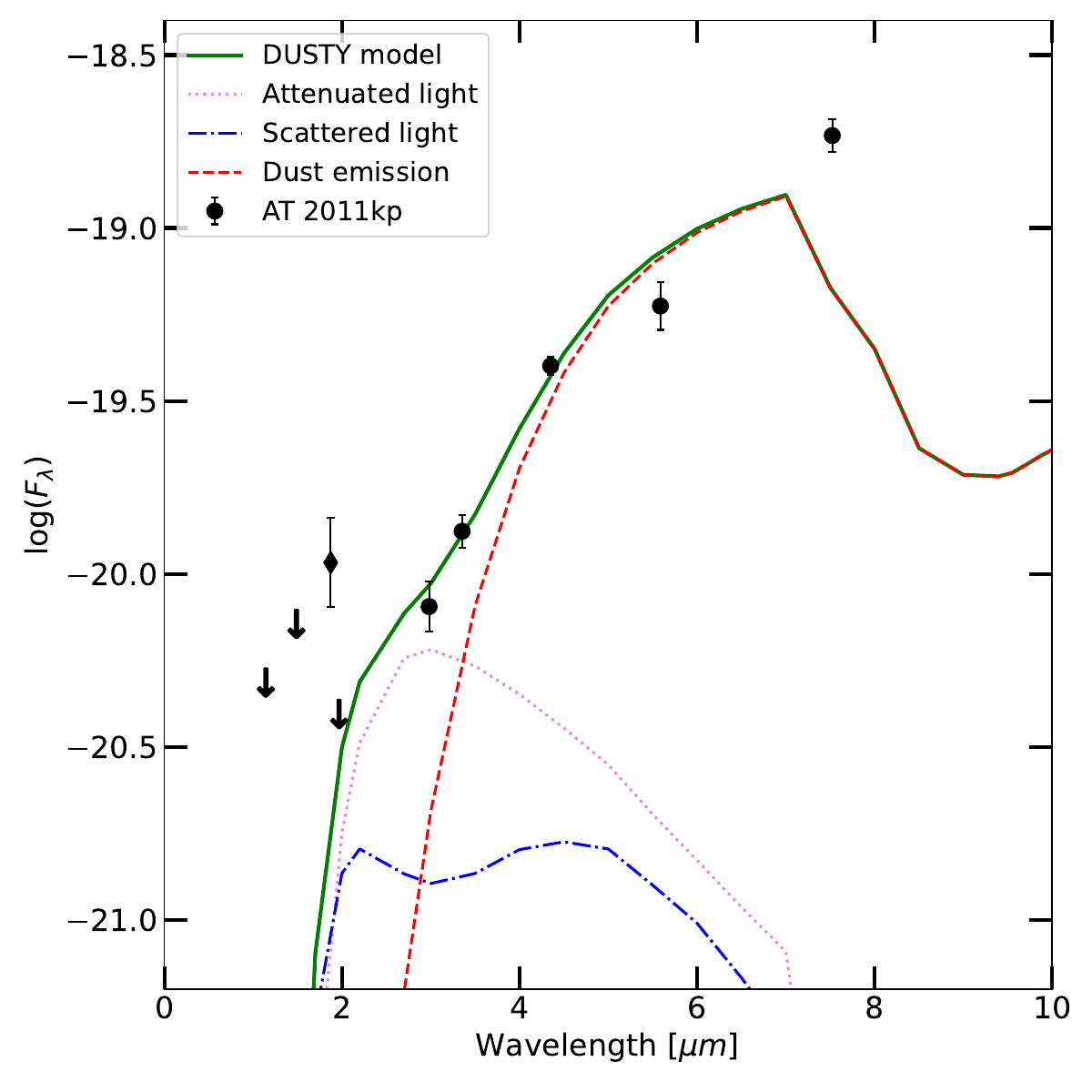}
\includegraphics[width=.78\columnwidth]{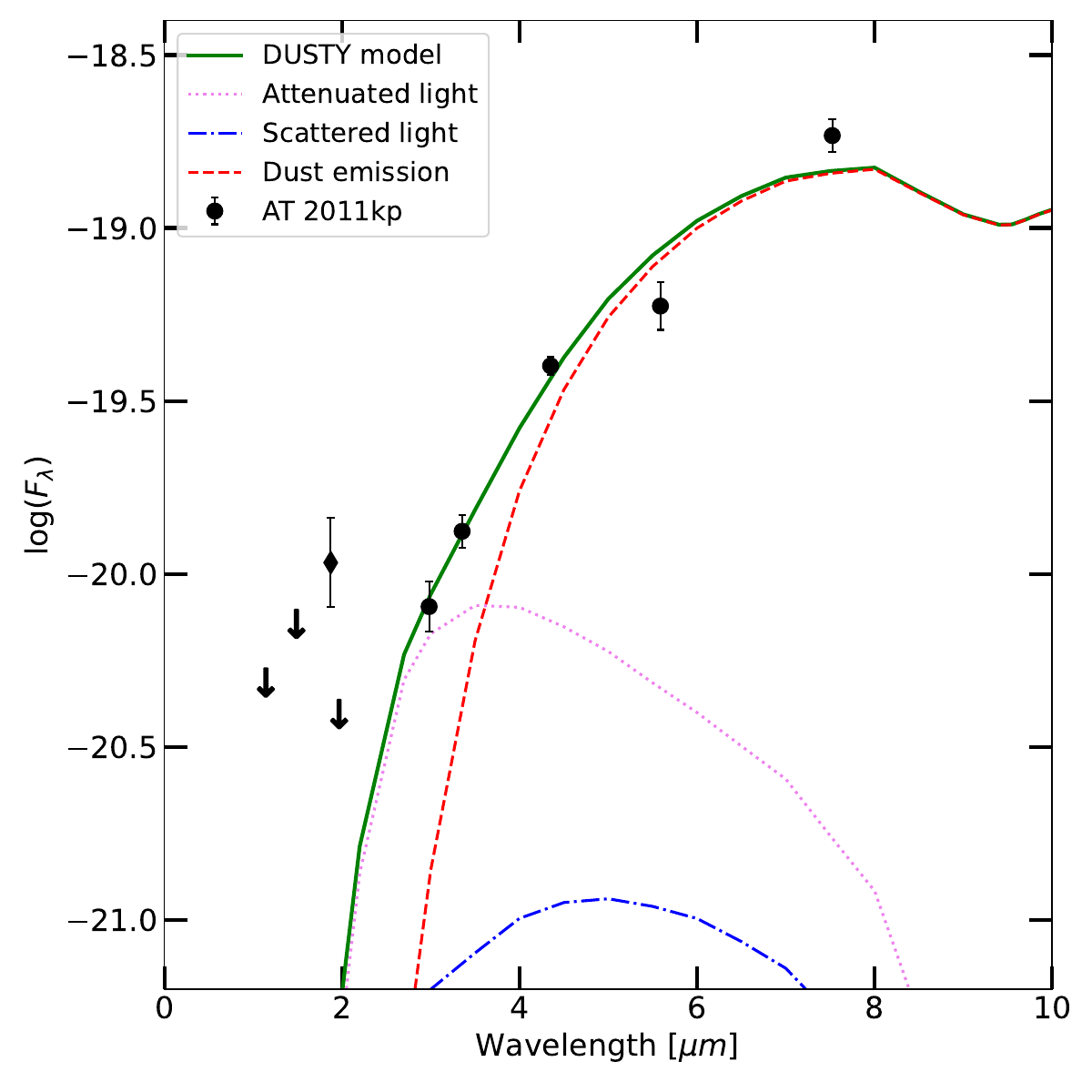}
\includegraphics[width=.78\columnwidth]{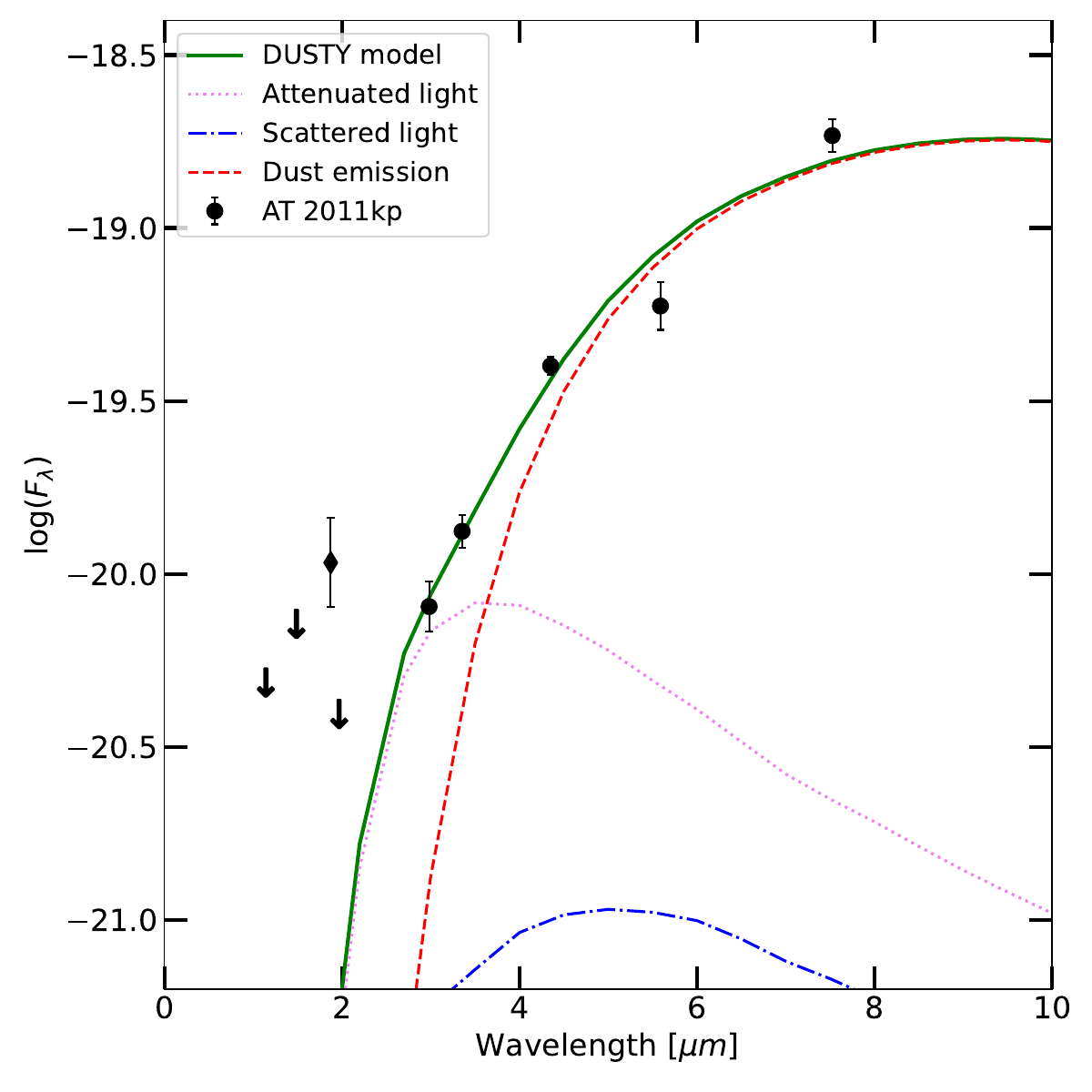}
\caption{Spectral energy distribution of AT 2011kp as obtained from JWST observations performed in 2023 and 2024 (13 years after the LRN onset), from 1.15 to 7.7 $\mu$m (black points). Downwards arrows mark upper limits. 
The SED is best-fitted with three \texttt{DUSTY} models with different chemical compositions: 100\% silicates (top panel), 50\% silicates + 50\% AMC (middle panel), and 100\% AMC (bottom panel).
Contributions from dust emission, attenuated, and scattered light are also shown.
The excess at 1.8 $\mu$m is marked with a rhombus.
}
\label{fig:SED_NGC4490OT}
\end{figure}

The origin of the excess flux at 7.7 $\mu$m could be due to the presence of dust composed of silicates - indicative of an O-rich environment, which have a peak in opacity at around 8-10 $\mu$m \citep{Draine1984ApJ...285...89D}.
Alternatively (though less likely), the $F770W$-filter excess can also be caused by emission from polycyclic aromatic hydrocarbons (PAHs), observed in the MIR SED and spectra of other IGTs \citep{Prieto2009ApJ...705.1425P, Valerin2025A&A...695A..42V}, which instead are typical of C-rich ambients.
Observations of the AT~2011kp field with JWST/MIRI at longer wavelengths (with $F1000W$ and even redder filters, and/or a MIR spectrum) would help resolving this conundrum.

The most important outcome of our analysis is that a significant amount of dust has likely formed after the LRN event, and has progressively cooled over the course of the past decade.
To understand if LRNe can be considered dust factories (as suggested also by \citealt{Karambelkar2025arXiv250803932K}), it is interesting to compare the results of the late-time follow-up of LRN AT 2011kp with those of two ILRTs with extensive long-term monitoring in the IR domain, viz. AT 2008jd (best known as NGC300-2008OT, e.g. \citealt{Ohsawa2010ApJ...718.1456O, Kochanek2011ApJ...741...37K}) and AT 2019abn \citep{Valerin2025A&A...695A..42V}. The two objects were observed by WISE, during the Cold mission (also including the $W3$ and $W4$ filters at 12 and 22 $\mu$m), and JWST/NIRCam and MIRI, respectively.
A dust temperature of 580 K and 380 K was inferred for AT 2008jd (at phase +2 years) and AT 2019abn (after 5 yrs, although influenced by the presence of emission from PAH), respectively \citep{Valerin2025A&A...695A..42V}. From this comparison, the dust temperature seems to fade more rapidly in ILRTs than in LRNe. 
In brief, these two types of IGTs have different properties during the explosive event in the optical domain whilst, for both, the SEDs at late epochs peak in the MIR region, although with different timescales. This is another parameter that may support the observational discrimination between the two IGT families. It also suggests that both classes are efficient dust producers.

\subsection{Dust modelling and masses}\label{sect:dust_masses}
Assuming that the source of the MIR radiation is heated dust, we can estimate its mass, temperature and chemical composition by fitting the IR SED with dust emission models.
We used the radiative transfer code \texttt{DUSTY} \citep{Ivezic1997MNRAS.287..799I, Ivezic1999astro.ph.10475I, Ivezic1999ascl.soft11001I} to derive these physical parameters. We assume a simple scenario with a spherically symmetric dust shell and a central hot source, a standard \cite{Mathis1977ApJ...217..425M} distribution for dust grain sizes, a $\rho \propto r^{-2}$ radial density profile and a shell's relative thickness $Y=R_{\mathrm{out}}/R_{\mathrm{in}}=2$, as in \cite{Kochanek2012ApJ...759...20K}.
To estimate the best-fit model and the associated uncertainties, we explored the parameters space by implementing a grid of models as described in Appendix \ref{Dusty}, and select the model with the lowest $\chi^2$. For the central source, we assume a BB radiation with a temperature of 3500 K, close to that estimated for the stellar remnants of LRNe in Sect. \ref{sect:SEDs}. 
We considered multiple combinations of chemical compositions of the dust, for both the cases of O-rich (in the form of silicates) and C-rich ambients. We prefer amorphous carbon (AMC) instead of graphite as in C-rich environments dust tends to be in the form of the former \citep{Zubko1996MNRAS.282.1321Z, Marini2021A&A...647A..69M}.
Anyway, the environment around an M-type star and a merger event is more likely to be O-rich rather than C-rich, and therefore we expect the dust to be composed of silicates \citep{Banerjee2004ApJ...615L..53B}.
The mass of the dust is determined through the same procedure as in \cite{Lau2025ApJ...983...87L} (see the equations in their Appendix A). The parameters associated to each best-fit model are presented in Table \ref{Tab:dust}.

To characterise the IR emission of AT~1997bs at late phases, we start reanalysing the SST observations obtained in 2004 and presented by \citet{Kochanek2012ApJ...758..142K} and \citet{VanDyk2012ApJ...746..179V}. Our analysis suggests that a source is visible in all four IRAC filters. With these data, we construct the MIR SED about seven years after the discovery of AT 1997bs. 

For AT 2011kp, the best-model fits are found with a mixed composition of both silicates and AMC, and with pure AMC (Fig. \ref{fig:SED_NGC4490OT}). The optical depth $\tau_V$ of the shell is 46 and 33, respectively. The temperature of the dust at the inner side of the shell is 450-475 K, while the inner radii of the dusty shell are also similar to each other, at 370 and 470 AU, respectively. Finally, we derive dust masses of 4.6 and 2.0 $\times 10^{-4}$ \Msun.
The velocity needed by the ejecta to reach a distance of 370 (470) AU in 12.5 years is $\sim$140 (180) \kms, the same determined by \cite{Smith2016} (160 \kms).
These models fit reasonably well the data obtained at wavelengths longer than 3 $\mu$m, including the excess at 7.7 $\mu$m. However, they miss the points at shorter ones. In particular, a strong excess in the $F187N$ filter remains, which could be explained with a Pa$\alpha$ emission line from the circumstellar medium (CSM), which suffers less extinction than the remnant.
The model with a 100\% silicates composition has a much larger $\chi^2$ and is thus unfavoured.

For AT 1997bs, we changed the source temperature to 4000~K, equal to that of the remnant observed 20 years later.
A best-fit (Fig. \ref{fig:SED_97bs}, middle panel) is found for a model of a dusty shell composed entirely of silicates, at a temperature on the inner side of 600 K, and an optical depth $\tau_V$ of $\sim$68. The dusty shell is located 300 AU from the central source, and contains $1.0 \times 10^{-3}$~\Msun\ of dust. 
Given the shorter amount of time passed since the merger, at this stage the dust is hotter with respect to AT~2011kp.
Another fit, this time with a shell made only of AMC, is found for a model again with $T_{\rm dust,in}=600$ K, and $\tau_V=$16 (Fig. \ref{fig:SED_97bs}, bottom panel). The inner radius of the dusty shell is 320 AU, and amasses $7.0 \times 10^{-5}$ \Msun\ of dust. 
The lower mass of dust in the case of dust made of AMC is due to its opacity per unit mass being higher than for silicates, and an overall lower $\tau_V$.
These distances (300-320 AU) travelled in 7 years correspond to an ejecta velocity of $\sim$200 \kms.

As a reference, the mass of dust produced by V838~Mon is also a few $\times 10^{-4}$~\Msun\,\citep{Kimeswenger2002MNRAS.336L..43K, Banerjee2003ApJ...598L..31B, Exter2016A&A...596A..96E}.
Assuming a standard gas-to-dust ratio of 100:1 (as in \citealt{Dwek1983ApJ...274..175D, Mauerhan2018MNRAS.473.3765M}, but see \citealt{Sarangi2025ApJ...993...94S}), the dust masses we inferred imply a total CSM mass of few $10^{-2}$ \Msun, which can also explain the putative detection of the Pa$\alpha$ emission line in the late-time SED of AT 2011kp.

The dust masses of AT 1997bs and AT 2011kp are similar to those found for other LRNe at very late times by \cite{Karambelkar2025arXiv250803932K}, and higher than the range of masses determined for AT 2014ib ($3\times10^{-6}$ to $3\times10^{-5}$~\Msun) by \cite{Mauerhan2018MNRAS.473.3765M}. 
However, these values are lower than those found for other types of IGTs, such as ILRTs (up to $2\times10^{-3}$~\Msun, \citealt{Valerin2025A&A...695A..42V}) and in normal SNe \citep[e.g.][]{Fabbri2011MNRAS.418.1285F, Bevan2016MNRAS.456.1269B, Bevan2020ApJ...894..111B}\footnote{\url{https://nebulousresearch.org/dustmasses/}}.

As a final note, the dust masses we obtained are likely to be lower limits because our longest-wavelength observations are at around 8 $\mu$m. Therefore, we miss the contribution from much colder, further away dust that may be present.

\begin{table*}\centering
\caption{Properties of the best-fit DUSTY models to the SEDs of AT 2011kp and AT 1997bs.
}
\label{Tab:dust}
\begin{threeparttable}
\begin{tabular}{ccccccccc}
\hline
\multicolumn{9}{c}{AT 2011kp (+12.5 years)}\\
\hline
$T_{\rm s}$ (K) & $T_{\rm d}$ (K) & composition & $\chi^2$ & $\tau_V$ & $R_{\rm dust, in}$ (cm)\tnote{1} & $R_{\rm source}$ (cm)\tnote{1} & $L_{\rm bol}$ (erg/s) & $M_{\rm dust}$ (\Msun) \\
\hline
3500 & 450 & 100\% sil & 42.7 & 133.5$\pm$9.5 & 4.35e+15 & 3.81e+13 & 1.56e+38 & (1.8$\pm$0.5)$\times10^{-3}$ \\
3500 & 475 & 50\% sil+50\% AMC & 13.2 & 46.0$\pm$2.6 & 5.56e+15 & 5.15e+13 & 2.80e+38 & (4.6$\pm$1.3)$\times10^{-4}$ \\
3500 & 450 & 100\% AMC & 11.5 & 33.0$\pm$2.3 & 7.08e+15 & 4.88e+13 & 2.55e+38 & (2.0$\pm$0.6)$\times10^{-4}$ \\
\hline
\multicolumn{9}{c}{AT 1997bs (+7 years)}\\
\hline
4000 & 600 & 100\% sil & 0.05 & 68.5$\pm$18.0 & 4.51e+15 & 5.64e+13 & 5.82e+38 & (1.0$\pm$0.3)$\times10^{-3}$ \\
4000 & 600 & 100\% AMC & 0.31 & 16.0$\pm$2.9 & 4.85e+15 & 5.09e+13 & 4.75e+38 & (6.9$\pm$1.3)$\times10^{-5}$ \\
\hline
\end{tabular}
    \begin{tablenotes}\footnotesize
    \item[1] The error on the radii is of the order of 1\%, plus the systematic uncertainty on the DM.
    \end{tablenotes}
\end{threeparttable}
\end{table*}

\subsection{Concluding remarks}
In this work, we have inspected very late optical, NIR and MIR images for a sample of LRNe studied in the literature.
The main goal was the characterisation of the LRN survivors. The sample we used includes two IGTs (AT 2007sv and AT 2015fx) that were previously classified as SN impostors, produced by bright outbursts from massive stars. Re-analysis of their previously observed properties together with the new argument we offer of late detection in the NIR support the reclassification of the two events as LRNe.
This reclassification is a first step in our effort to simplify the observational `zoo' of IGTs, which contains a significant variety. 
In fact, we propose that individual short-duration eruptive events are more likely to result from merging events in close binaries rather than the violent eruptive mass loss attributed to very massive stars.

We recovered AT 2011kp 13 years after the LRN event in JWST images, demonstrating that LRNe may remain bright in the MIR for many years.
The SED of AT 2011kp in 2024 can be explained by a cool remnant source surrounded by a cold dusty cocoon at a temperature of $\sim$450-475 K plus emission from the H Pa$\alpha$ line.
Between 3 and 6 years after the LRN onset, the SST light curves of AT~2011kp were declining in luminosity without becoming redder, as one would expect from a cooling object. However, the BB radius remained roughly constant, while the BB temperature decreased as $T \propto t^{-1/2}$, which are signatures of an IR echo propagating outwards through a dusty shell.
After more than six years, the dusty circumstellar shell had likely fragmented in a similar way as directly observed for V838~Mon through resolved HST images only a few months after the event \citep{Bond2003Natur.422..405B}. This fragmentation allowed us to peer through the CSM and see the product of the merger. 
At these very late epochs, the IR echo started to disappear, while the dust shell cooled further. Therefore, in the next decades, only the central merger will remain observable, with properties that are expected to be similar to those currently observed in AT~1997bs.

JWST detected AT 1997bs in the NIR domain nearly 30 years after its outburst, but no source has been detected in the MIR.
Its SED points towards a surviving red and expanded star as the outcome of the common-envelope ejection and the subsequent binary coalescence. Dust was surely present and contributed to the IR emission in 2004, seven years after the outburst, but was no longer producing MIR emission at the time of the 2023-2024 observations, hence unveiling the surviving object. Another possibility is that in the 20 years since the SST observation, the dust might have cooled down to the point that it is now emitting in the far-IR, hence impeding us from detecting it.
The most recent observations reveal the red survivor of AT~1997bs, whose luminosity remained constant in the $F814W$ filter between 2014 and 2023, supporting the idea of a fragmented CSM as seen in V838 Mon \citep{Bond2003Natur.422..405B}. 

The remnant of AT 2019zhd, a LRN from a lower mass progenitor, is also compatible with a very cool and inflated red supergiant, similar to that found for OGLE-2002-BLG-360 \citep{Steinmetz2025A&A...699A.316S}.
Therefore, three LRNe were observed to end up as such up to now.

In conclusion, LRNe remain bright and visible in the MIR for years after coalescence due to persistent emission from newly formed dust. From their IR SEDs, we inferred dust masses of the order of $10^{-4}$ \Msun.
Lastly, long after the merger, a cool and expanded remnant becomes discernible, similar to a red giant or supergiant star, demonstrating that these transients are not terminal events.

\section*{Data availability}
All data analysed in this paper are publicly available through dedicated databases, such as IRSA and MAST, and can be obtained from \url{https://dx.doi.org/10.17909/ky92-0c89}.
All the photometric measurements are tabulated in the photometry tables, which are available in electronic form at the CDS.

\begin{acknowledgements}
\begin{small}
We thank the anonymous referee for the constructive comments which helped to improve the quality of the manuscript.

AR acknowledges financial support from the GRAWITA Large Program Grant (PI P. D’Avanzo). AR, AP, GV acknowledge financial support from the PRIN-INAF 2022 `Shedding light on the nature of gap transients: from the observations to the models'.

This work is based on observations made with the NASA/ESA/CSA James Webb Space Telescope. The data were obtained from the Mikulski Archive for Space Telescopes at the Space Telescope Science Institute, which is operated by the Association of Universities for Research in Astronomy, Inc., under NASA contract NAS 5-03127. These observations are associated with programs ID 1783, 2107, 4793.

This research is based on observations made with the NASA/ESA Hubble Space Telescope obtained from the Space Telescope Science Institute, which is operated by the Association of Universities for Research in Astronomy, Inc., under NASA contract NAS 5–26555. These observations are associated with programs ID 13364, 13442, 13773, 14668, 15645, 16239, 16800, 16801, 17070.

This work is based in part on observations made with the Spitzer Space Telescope, which was operated by the Jet Propulsion Laboratory, California Institute of Technology under a contract with NASA.

This publication makes use of data products from NEOWISE, which is a joint project of the Jet Propulsion Laboratory/California Institute of Technology and the University of Arizona. NEOWISE is funded by NASA.
\end{small}
\end{acknowledgements}

\bibliographystyle{aa}
\bibliography{bib}

\begin{appendix}

\section{Additional plots}\label{Appendix}

\begin{figure}[ht]
\includegraphics[width=1\columnwidth]{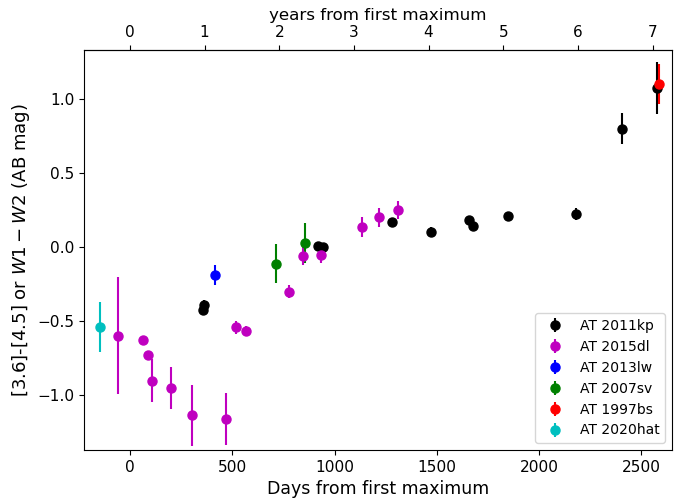}
\caption{Mid-infrared [3.6]-[4.5] or $W1-W2$ colour curves of AT~2011kp, AT 2015dl, AT 2013lw, AT 2007sv, AT 1997bs, AT~2020hat between 5 months before and 7 years after their respective first maximum. 
The objects show a remarkably similar evolution.}
\label{fig:MIR_color}
\end{figure}

\begin{figure}[h]
\includegraphics[width=0.95\columnwidth]{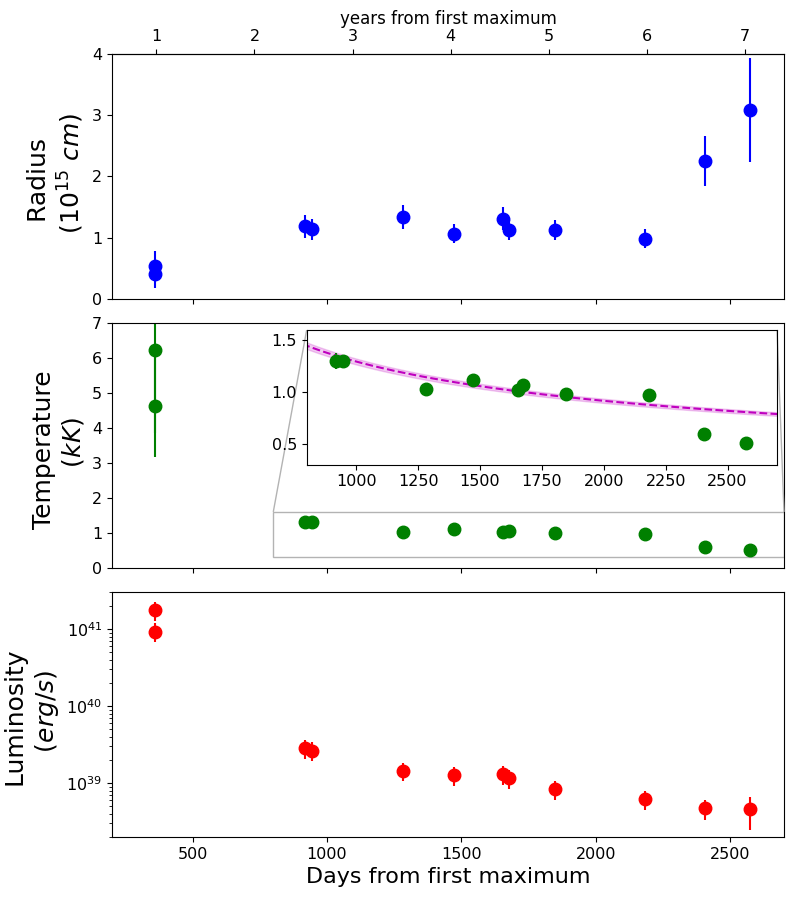}
\caption{Evolution of the BB radius (top), temperature (middle) and the bolometric luminosity (bottom) of AT 2011kp, estimated through a fit of a Planckian function on the SST $Ch1$ and $Ch2$ data. In the middle panel, the late-time temperature evolution is zoomed-in, together with the best-fit $At^{-0.5}$ function on the data between +2 and +6 years, showing a good match.}
\label{fig:RTL_11kp}
\end{figure}

\section{\texttt{DUSTY} best-fit models determination}\label{Dusty}

\begin{figure}[h]
\includegraphics[width=0.9\columnwidth]{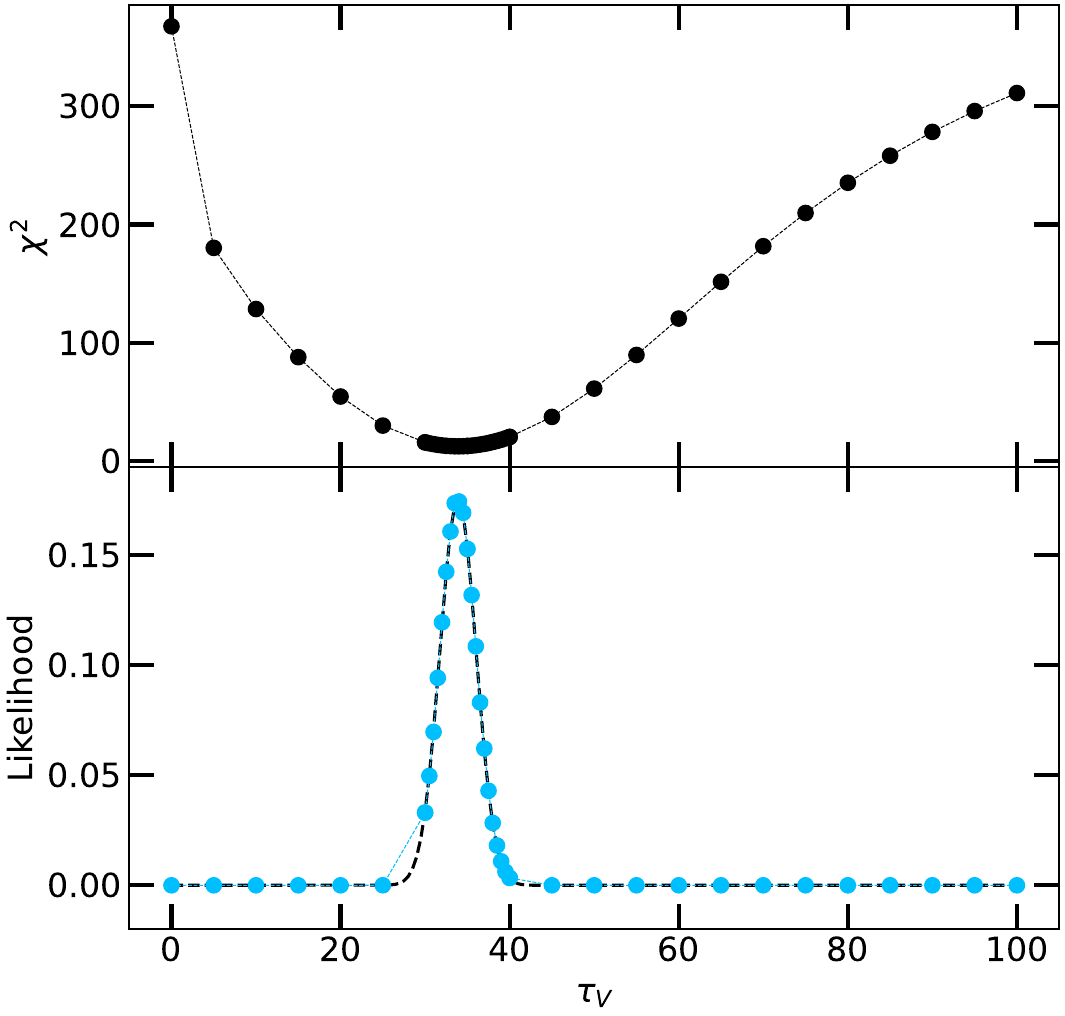}
\caption{Plot of $\chi^{2}$ and normalised likelihood for 42 DUSTY models generated with the same initial parameters while only varying the optical depth $\tau_V$.}
\label{fig:Tau_V}
\end{figure}

In order to identify the \texttt{DUSTY} model that best fits our observations, we constructed a grid of models in subsequent steps. Firstly, once fixed the temperature of the central source (T$_{s}$), the density profile of the dust ($k$) and the relative width of the dust shell ($Y$), we also fixed the dust temperature at the inner radius (T$_{d}$) and produced 21 \texttt{DUSTY} models with optical depths in the $V$-band ($\tau_{V}$) spanning from 0 to 100 with a step of five. When necessary, for example with some silicate dust configurations, we extended the grid to larger optical depths (typically up to 200). We introduced a multiplicative factor as a free parameter to scale each model to the observations (physically, this is a proxy for the dust radius, which is not needed for generating \texttt{DUSTY} models), obtaining a $\chi^{2}$ value for each model. We then refined the grid around the model with the minimum value of $\chi^{2}$, generating an additional set of 21 \texttt{DUSTY} models with optical depth $\tau_{V}$ centred on the value of $\tau_{V}$ providing the best $\chi^{2}$, this time with a step of 0.5. Again, thanks to the free multiplicative parameter we are able to fit the models to the observations and determine which $\tau_{V}$ provides the best results for that specific set of parameters. An example of the resulting $\chi^{2}$ and normalised likelihoods for the specific case of T$_{s}$ = 3500 K, T$_{d}$ = 450 K, $k$ = 2, $Y$ = 2 and AMC composition can be seen in Figure \ref{fig:Tau_V}. In the lower panel, the likelihood is calculated from the $\chi^{2}$ value as

\begin{equation}
\hspace{2.5cm} L(\tau) = exp \left( \frac{-(\chi^{2}(\tau)-\chi^{2}_{min})}{2}) \right) ,
\end{equation} 

and we normalised it so that its integral over the considered interval is equal to one. By fitting a Gaussian function to the likelihood values, we are able to associate an uncertainty to $\tau_{V}$ of the best model for each set of input parameters, by taking its standard deviation $\sigma$.

\begin{figure}
\includegraphics[width=0.8\columnwidth]{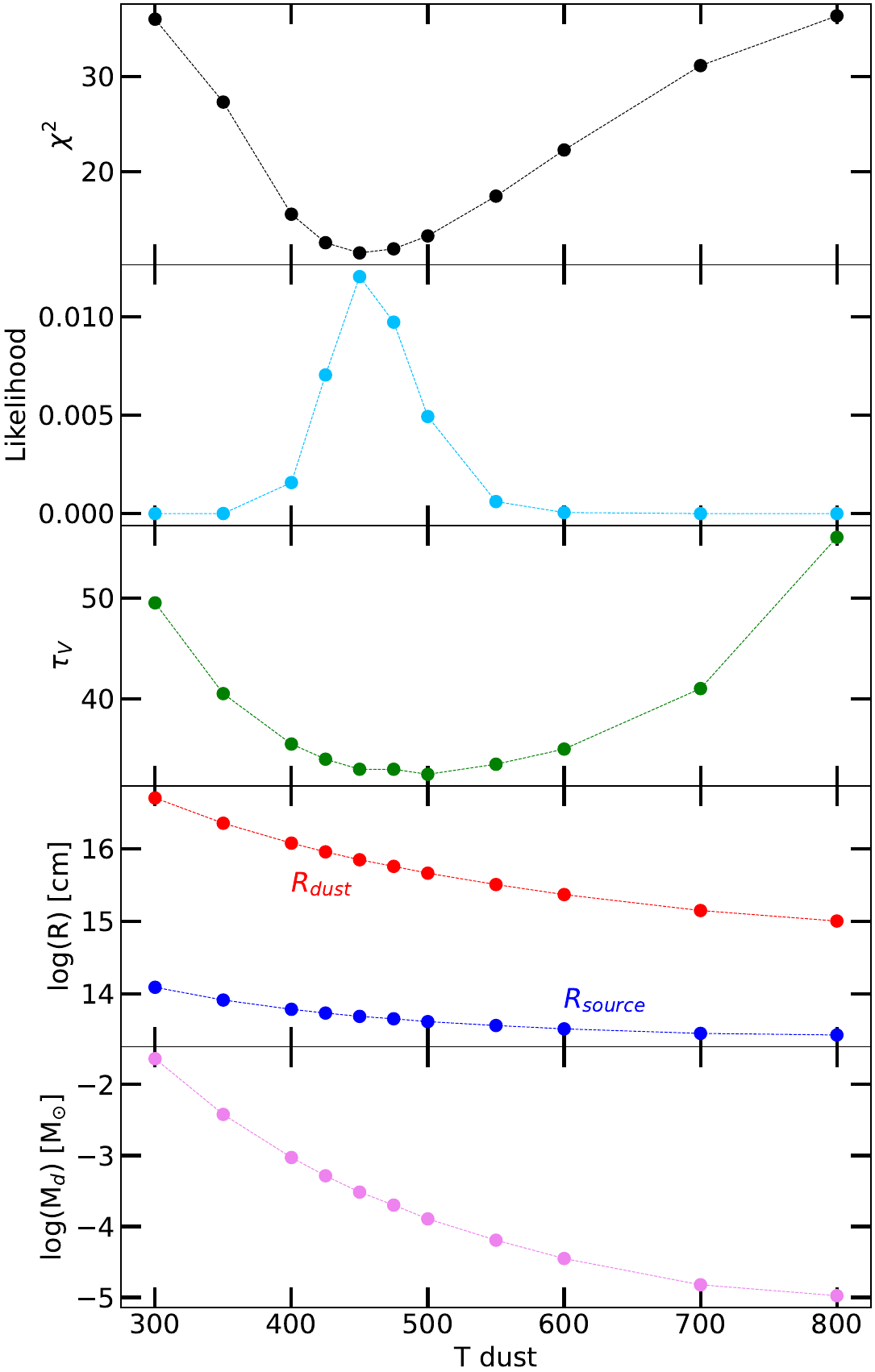}
\caption{Parameters of the best models against each considered dust temperature T$_{d}$. From top to bottom panels: $\chi^{2}$ and likelihood parameters, optical depth $\tau_{V}$, dust and source radii, and dust mass.}
\label{fig:results}
\end{figure}

We then repeated this process for a set of inner dust temperatures T$_{d}$, initially with a step of 100 K, and then refined the grid with steps of 50 K (and for AT 2011kp down to 25 K) around the best solution found. In Figure \ref{fig:results} we display the best models obtained, resulting from minimising the $\chi^{2}$ of 42 models with varying $\tau_V$ for each T$_{d}$ considered. The fixed parameters for the models shown in Figure \ref{fig:results} are T$_{s}$ = 3500 K, $k$ = 2, $Y$ = 2 and AMC composition. By observing the distribution of $\chi^{2}$ and likelihood, it is clear that there is a preferred inner dust temperature for these fixed parameters, which is therefore selected as the best model for this given configuration. The last three panels of Figure \ref{fig:results} show the variations of the optical depth in the $V$-band $\tau_{V}$, dust radius, source radius, and dust mass for the best models at each considered value of T$_{d}$.
All the steps described above were repeated for a composition of pure silicates and, in the case of AT 2011kp, also for a mixture of silicates and AMC.

In order to explore the effects of changing the properties of the central source, we also performed the procedure described above increasing the temperature of the central star to 7000 K. We find that similar models are obtained in terms of quality of the fits with limited changes to the dust temperature ($\sim$20\%); however, especially for AMC models, the dust radius tends to increase with increasing temperature of the central source (up to a factor of two), as the optical depth diminishes, with competing effects on the dust mass, which indeed is not strongly affected by the changes of the central source. The parameter most affected is the radius of the central source, which shrinks significantly. In the considered cases, with the available data, it is challenging to securely infer the properties of the central source from the observed data.

\end{appendix}
\end{document}